%% file: paper.tex
\documentclass[journal]{IEEEtran}
\newif\iffullpaper
\fullpapertrue

\usepackage[keeplastbox]{flushend}
\usepackage{cite}
\usepackage[subrefformat=parens, labelformat=parens]{subfig}
\usepackage[font=small]{caption}
\usepackage{times}
\usepackage{siunitx} 
\usepackage{color}
\usepackage{lipsum}
\usepackage{algorithm, algorithmicx, algpseudocode}
\usepackage{amsmath, amssymb, amsthm} 
\usepackage[shortlabels]{enumitem}
\usepackage{upgreek}
\usepackage{bm, bbold, dsfont}
\usepackage{tabularx}
\usepackage{mathtools}
\usepackage[colorinlistoftodos]{todonotes}
\usepackage{booktabs}

\newcommand{\GZcomment}[1]{{\textbf{\color{red}$<$GZ$>$: #1.}}}

\newcommand{\littlesum}{\mathop{\textstyle\sum}}
\newcommand{\littleint}{\mathop{\textstyle\int}}

\usepackage{graphicx}
\usepackage{epstopdf}

\newtheorem{theorem}{Theorem}

\newtheorem{definition}{Definition}

\newtheorem{mylemma}{Lemma}

\newtheorem{corollary}{Corollary}

\newtheorem{proposition}{Proposition}

\newtheorem{remark}{Remark}

\usepackage[colorinlistoftodos]{todonotes}

\ifCLASSINFOpdf
\else
\fi
\usepackage{url}
\begin{document}

\setlength{\abovedisplayskip}{3pt} 
\setlength{\belowdisplayskip}{3pt} 

\sloppy

\title{Hybrid Scheduling in Heterogeneous \\ Half- and Full-Duplex Wireless Networks}

\author{Tingjun~Chen, 
Jelena~Diakonikolas, 
Javad~Ghaderi, 
and~Gil~Zussman
\thanks{This research was supported in part by NSF grants ECCS-1547406, CNS-1650685, and CNS-1717867, ARO grant 9W911NF-16-1-0259. A partial and preliminary version of this paper appeared in IEEE INFOCOM'18, Apr. 2018~\cite{chen2018hybrid}, and some results were presented in the Asilomar Conference on Signals, Systems, and Computers (invited), Oct. 2018.}
\thanks{T. Chen, J. Ghaderi, and G. Zussman are with the Department of Electrical Engineering, Columbia University, New York, NY, USA (email: \{tingjun, jghaderi, gil\}@ee.columbia.edu).}
\thanks{J. Diakonikolas is with the Department of Statistics, University of California at Berkeley, Berkeley, CA, USA (e-mail: jelena.d@berkeley.edu).}
}

\maketitle

\input{tex/nomenclature}


\begin{abstract}
Full-duplex (FD) wireless is an attractive communication paradigm with high potential for improving network capacity and reducing delay in wireless networks. Despite significant progress on the physical layer development, the challenges associated with developing medium access control (MAC) protocols for \emph{heterogeneous} networks composed of both legacy half-duplex (HD) and emerging FD devices have not been fully addressed. Therefore, we focus on the design and performance evaluation of scheduling algorithms for infrastructure-based \emph{heterogeneous} HD-FD networks (composed of HD and FD users).
{We first show that \emph{centralized} Greedy Maximal Scheduling ({\GMS}) is throughput-optimal in heterogeneous HD-FD networks. We propose the \emph{Hybrid}-{\GMS} ({\HGMS}) algorithm, a distributed implementation of {\GMS} that combines {\GMS} and a queue-based random-access mechanism.}
\emph{We prove that {\HGMS} is throughput-optimal}.
Moreover, we analyze the delay performance of {\HGMS} by deriving lower bounds on the average queue length. We further demonstrate the benefits of upgrading HD nodes to FD nodes in terms of throughput gains for individual nodes and the whole network. Finally, we evaluate the performance of {\HGMS} and its variants in terms of throughput, delay, and fairness between FD and HD users via extensive simulations. We show that in heterogeneous HD-FD networks, {\HGMS} achieves $16$--$30\times$ better delay performance and improves fairness between HD and FD users by up to $50\%$ compared with the fully decentralized {\QCSMA} algorithm.
\end{abstract}


%
\IEEEpeerreviewmaketitle

\section{Introduction}\label{sec:intro}
\input{tex/intro}

\section{Related Work}\label{sec:related}
\input{tex/related}

\section{Model and Preliminaries}
\label{sec:model}
\input{tex/model}

\section{Scheduling Algorithms and Main Result}
\label{sec:sched-alg}
\input{tex/sched_alg}

\section{Proof of Theorem~\ref{thm:stability} via Fluid Limits}
\label{sec:mainproof}
\input{tex/proof_stability}

\section{Lower Bounds on the Average Queue Length}
\label{sec:delay}
\input{tex/delay}

\section{Benefits of Introducing FD-Capable Nodes}
\label{sec:benefits}
\input{tex/benefits}

\section{Simulation Results}
\label{sec:sim}
\input{tex/simulation}

\section{Conclusion}
\label{sec:conclusion}
We presented a hybrid scheduling algorithm, {\HGMS}, for heterogeneous HD-FD infrastructure-based networks. {\HGMS} is distributed at the users and leverages different degrees of centralization at the AP to achieve good delay performance while being provably throughput-optimal. {We also derived lower bounds on the average queue length to evaluate the delay performance of {\HGMS}.} We further illustrated various aspects of the performance of {\HGMS} and compared it to the classical {\QCSMA} through extensive simulations. We also illustrated benefits and fairness-efficiency tradeoffs arising from incorporating FD users into existing HD networks. There are several important directions for future work. We plan to expand the results to multi-channel networks with general topologies and to study the impact of imperfect SIC on the scheduling algorithms and their performance. In addition, an experimental evaluation of {\HGMS} on a real wireless testbed is an important step towards a provably-efficient and practical MAC layer for HD-FD networks.




\bibliographystyle{IEEEtran}
\bibliography{infocom18}

\iffullpaper

\appendices

\section{Proof of Proposition~\ref{prop:GMS}}
\label{append:GMS}
\input{tex/proof_OLoP}

\section{Proof of Lemma~\ref{lem:DTMC-steady-state}}
\label{append:DTMC-steady-state}
\input{tex/proof_steady_state}

\section{Proof of Lemma~\ref{lem:fluid-limit-eq}: Fluid limit Equations}
\label{append:fluid-limit-eq}
\input{tex/proof_fluid_limit_eq}

\section{Proof of Proposition~\ref{prop:delay-lb-hgms}}
\label{append:proof-delay-lb-hgms}
\input{tex/proof_delay_lb_hgms.tex}

\fi

\end{document}

%% file: tex/nomenclature.tex
\newcommand{\NumUserFD}{N_{F}}
\newcommand{\NumUserHD}{N_{H}}
\newcommand{\NumUser}{N}
\newcommand{\numUserFD}{N_{F}}
\newcommand{\numUserHD}{N_{H}}
\newcommand{\numUser}{N}

\newcommand{\setNode}{\mathcal{V}}
\newcommand{\setUser}{\mathcal{N}}
\newcommand{\setUserFD}{\setUser_{F}}
\newcommand{\setUserHD}{\setUser_{H}}

\newcommand{\setULDL}{\{\textrm{u}, \textrm{d}\}}

\newcommand{\setLink}{\mathcal{E}}
\newcommand{\setLinkFD}{\setLink_{F}}
\newcommand{\setLinkHD}{\setLink_{H}}
\newcommand{\setLinkMax}{\mathcal{E}_{\textrm{max}}}
\newcommand{\setLinkMin}{\mathcal{E}_{\textrm{min}}}

\newcommand{\nodeUL}[1]{v_{#1}^{\textrm{u}}}
\newcommand{\nodeDL}[1]{v_{#1}^{\textrm{d}}}

\newcommand{\Tput}{S}
\newcommand{\tput}{S}

\newcommand{\myProb}[1]{\mathbb{P}\left\{#1\right\}}
\newcommand{\myExp}[1]{e^{#1}}
\newcommand{\myE}[1]{\mathbb{E}[#1]}
\newcommand{\myTwoNorm}[1]{||#1||}
\newcommand{\myVar}[1]{\mathrm{Var}\left[#1\right]}
\newcommand{\myInv}[1]{{#1}^{-1}}

\newcommand{\arr}{A}
\newcommand{\arrLink}[1]{\arr_{#1}}
\newcommand{\arrVec}{\mathbf{\arr}}
\newcommand{\arrUL}[1]{\arr_{#1}^{\textrm{u}}}
\newcommand{\arrDL}[1]{\arr_{#1}^{\textrm{d}}}

\newcommand{\arrRate}{\lambda}
\newcommand{\arrRateLink}[1]{\arrRate_{#1}}
\newcommand{\arrRateVec}{\bm{\uplambda}}
\newcommand{\arrRateUL}[1]{\arrRate_{#1}^{\textrm{u}}}
\newcommand{\arrRateDL}[1]{\arrRate_{#1}^{\textrm{d}}}
\newcommand{\arrRateSum}[1]{\lambda_{#1}}

\newcommand{\serRate}{\mu}
\newcommand{\serRateVec}{\bm{\upmu}}
\newcommand{\serRateUL}[1]{\serRate_{#1}^{\textrm{u}}}
\newcommand{\serRateDL}[1]{\serRate_{#1}^{\textrm{d}}}

\newcommand{\isFDPair}{\land}
\newcommand{\notFDPair}{\lor}

\newcommand{\link}{l}
\newcommand{\linkFD}{\link_{f}}
\newcommand{\linkHD}{\link_{h}}
\newcommand{\linkUL}[1]{\link_{#1}^{\textrm{u}}}
\newcommand{\linkDL}[1]{\link_{#1}^{\textrm{d}}}

\newcommand{\UL}[1]{\link_{#1}^{\textrm{u}}}
\newcommand{\DL}[1]{\link_{#1}^{\textrm{d}}}
\newcommand{\setConflict}[1]{\mathcal{C}(#1)}

\newcommand{\queue}{Q}
\newcommand{\queueLink}[1]{\queue_{#1}}
\newcommand{\queueVec}{\mathbf{\queue}}
\newcommand{\queueUL}[1]{\queue_{#1}^{\textrm{u}}}
\newcommand{\queueDL}[1]{\queue_{#1}^{\textrm{d}}}

\newcommand{\queueEst}{\widetilde{\queue}}
\newcommand{\queueEstLink}[1]{\queueEst_{#1}}

\newcommand{\queueAvg}{\overline{\queue}}
\newcommand{\queueSum}{\mathcal{\queue}}

\newcommand{\sched}{X}
\newcommand{\schedLink}[1]{\sched_{#1}}
\newcommand{\schedUL}[1]{\sched_{#1}^{\textrm{u}}}
\newcommand{\schedDL}[1]{\sched_{#1}^{\textrm{d}}}
\newcommand{\schedVec}{\mathbf{\sched}}
\newcommand{\schedSub}{z}
\newcommand{\schedSubVec}{\mathbf{\schedSub}}

\newcommand{\setFeasSched}{\mathcal{S}}
\newcommand{\setDataSched}{\setFeasSched_{0}}

\newcommand{\schedData}{\schedVec}
\newcommand{\schedDataUL}[1]{\schedData_{#1}^{\textrm{u}}}
\newcommand{\schedDataDL}[1]{\schedData_{#1}^{\textrm{d}}}
\newcommand{\schedSymFDBD}[1]{\mathbf{f}_{#1}}
\newcommand{\schedSymHD}[1]{\mathbf{h}_{#1}}
\newcommand{\schedSymHDUL}[1]{\mathbf{h}_{#1}^{\textrm{u}}}
\newcommand{\schedSymHDDL}[1]{\mathbf{h}_{#1}^{\textrm{d}}}

\newcommand{\schedDec}{\mathbf{m}}
\newcommand{\schedDecUL}[1]{\schedDec_{#1}^{\textrm{u}}}
\newcommand{\schedDecDL}[1]{\schedDec_{#1}^{\textrm{d}}}

\newcommand{\probAcc}{\alpha}
\newcommand{\probAccVec}{\bm{\upalpha}}
\newcommand{\probAccLink}[1]{\probAcc_{#1}}

\newcommand{\optDL}{i^{\star}}

\newcommand{\probTX}{p}
\newcommand{\probTXSub}[1]{p_{#1}}
\newcommand{\probBarTXSub}[1]{\overline{p}_{#1}}
\newcommand{\probTXLink}[1]{p_{#1}}
\newcommand{\probBarTXLink}[1]{\overline{p}_{#1}}
\newcommand{\probTXUL}[1]{p_{#1}^{\textrm{u}}}
\newcommand{\probTXDL}[1]{p_{#1}^{\textrm{d}}}
\newcommand{\probBarTXUL}[1]{\overline{p}_{#1}^{\textrm{u}}}
\newcommand{\probBarTXDL}[1]{\overline{p}_{#1}^{\textrm{d}}}

\newcommand{\probSched}[1]{\pi(#1)}

\newcommand{\transProb}[2]{P({#1}, {#2})}
\newcommand{\transProbMat}{\mathbf{P}}

\newcommand{\capReg}{\Lambda}
\newcommand{\rateImprov}{\gamma}
\newcommand{\cvxHull}[1]{\mathrm{Co}(#1)}
\newcommand{\baseVec}[1]{\mathbf{e}_{#1}}
\newcommand{\zeroVec}{\mathbf{0}}
\newcommand{\oneVec}{\mathbf{1}}


\newcommand{\clique}{\mathcal{C}}

\newcommand{\MWS}{\textrm{MWS}}
\newcommand{\GMS}{\textrm{GMS}}
\newcommand{\QCSMA}{\textrm{Q-CSMA}}

\newcommand{\HGMS}{\textrm{H-GMS}}
\newcommand{\HGMSRDL}{\textrm{H-GMS-R}}
\newcommand{\HGMSLDL}{\textrm{H-GMS}}
\newcommand{\HGMSLDLE}{\textrm{H-GMS-E}}


%% file: tex/intro.tex
Full-duplex (FD) wireless -- an emerging wireless communication paradigm in which nodes can simultaneously transmit and receive on the same frequency -- has attracted significant attention~\cite{sabharwal2014band}. Recent work has demonstrated physical layer FD operation~\cite{duarte2012experiment, bharadia2013full,yang2015wideband,zhou2017integrated}, and therefore, the technology has the potential to increase network capacity and improve delay compared to legacy half-duplex (HD) networks. Based on the advances in integrated circuits-based implementations that can be employed in mobile nodes (e.g.,~\cite{yang2015wideband,krishnaswamy2016spectrum,zhou2017integrated,chen2019wideband}), we envision a gradual but steady replacement of existing HD nodes with the more advanced FD nodes. During this gradual penetration of FD technology, the medium access control (MAC) protocols will need to be carefully redesigned to not only support a \emph{heterogeneous} network of HD and FD nodes but also to guarantee fairness to the different node types.

Therefore, we focus on the design and performance evaluation of scheduling algorithms for heterogeneous HD-FD networks. In particular, we consider infrastructure-based random-access networks (e.g., IEEE 802.11) consisting of an FD access point (AP) and both HD and FD users in a single collision domain. Further, we consider a single channel which is shared by all the uplinks (ULs) and downlinks (DLs) between the AP and the users. To focus on fundamental limits due to the incorporation of FD nodes and to expose the main features of our scheduling algorithms, we assume perfect self-interference cancellation (SIC) at FD nodes. Yet, we expect that the results can be extended to more realistic settings by incorporating imperfect SIC.  

Traditionally, three approaches have been used for the design of wireless scheduling algorithms that can guarantee maximum throughput: 

\noindent\textbf{Maximum Weight Scheduling ({\MWS})}~\cite{tassiulas1992stability}, which relies on the queue length information and schedules non-conflicting links with the maximum total queue length. In contrast to the all-HD networks where only a single link can be scheduled at a time, in the considered setting the UL and the DL of any FD user can be scheduled simultaneously. Thus, to implement {\MWS}, queue length information needs to be shared between each FD user and the AP, which requires significant overhead. 

\noindent\textbf{Greedy Maximal Scheduling ({\GMS})}~\cite{dimakis2006sufficient}, which is a centralized policy that greedily selects the link with the longest queue, disregards all conflicting links, and repeats the process. Typically, {\GMS} has better delay performance than {\MWS} and {\QCSMA}. Although {\GMS} is equivalent to {\MWS} in an all-HD network, in general, it is not equivalent to {\MWS} and is not throughput-optimal in general topologies. 

\noindent\textbf{Queue-Length-based Random-Access Algorithms (e.g., {\QCSMA})}~\cite{ghaderi2010design,ni2012q}, which are fully distributed and do not require sharing of the queue length information between the users and the AP. These algorithms have been shown to achieve throughput optimality. However, they generally suffer from excessive queue lengths that lead to long delays.  

In this paper, we show that a combination of the two latter approaches guarantees maximum throughput and provides good delay performance in heterogeneous HD-FD networks. We first show by using the notion of Local Pooling~\cite{dimakis2006sufficient,birand2012analyzing} that {\GMS} is throughput-optimal in the considered HD-FD networks. However, since {\GMS} is fully centralized, we leverage ideas from distributed {\QCSMA} to develop the \textit{Hybrid}-{\GMS} ({\HGMS}) algorithm that combines centralized {\GMS} with distributed {\QCSMA}. The main feature of the proposed {\HGMS} algorithm is that instead of approximating {\MWS} (as done in ``traditional'' {\QCSMA}), it approximates {\GMS}.

The design of {\HGMS} leverages the fact that in infrastructure-based networks, the AP has access to all the DL queues and can resolve the contention among the DL queues (e.g., using longest-queue-first). In contrast, the users do not have access to all DL queues or to other UL queues, and therefore, must share the medium in a distributed manner, while ensuring FD operation when possible. 

We prove the throughput optimality of {\HGMS}  (namely, it can support any rate vector in the capacity region of heterogeneous HD-FD networks) by using the fluid limit technique. In contrast to the classical {\QCSMA}, the contention resolution of DL queues at the AP under the {\HGMS} algorithm can force a schedule that is \emph{not} with maximum weight (i.e., not {\MWS}). Hence, we make a connection to {\GMS} in fluid limits (which, as mentioned above, is throughput-optimal in heterogeneous HD-FD networks). We also present variants of {\HGMS} with different degrees of centralization. To understand the delay performance of {\HGMS}, in Section~\ref{sec:delay}, we derive two lower bounds on the average queue length: (i) a fundamental lower bound that is independent of the scheduling algorithm, and (ii) a stronger lower bound that takes into account the characteristics of the developed {\HGMS} and applies to all its non-adaptive variants. These lower bounds serve as benchmarks when evaluating the delay performance of {\HGMS}.

Before thoroughly evaluating {\HGMS} and its variants, we demonstrate the benefits of introducing FD-capable users into an all-HD network in terms of both network and individual throughput gains. Compared to the all-HD network, the considered heterogeneous HD-FD network can potentially double the  throughput for certain rate vectors within the capacity region, while the network throughput gain generally depends on both the number of FD users and the specific rate vector in which the network operates. Using simple examples, we show that when all links have equal rate, the throughput gain of the HD-FD network over the all-HD network increases with the number of FD users, and it reaches a gain of 2 when all users are FD-capable. We also demonstrate that it is generally possible for all users to experience improved individual throughput at the cost of lowering the priority of FD users, revealing an interesting \emph{fairness-efficiency tradeoff}.

Finally, we present extensive simulation results to evaluate the different variants of the {\HGMS} algorithm and compare them to the classical {\QCSMA} algorithm. We primarily focus on delay performance and fairness between FD and HD users, but also illustrate throughput gains. We consider a wide range of arrival rates and varying number of FD users. The results show that in heterogeneous HD-FD networks, {\HGMS} achieves $16$--$30\times$ better delay performance and improves fairness between HD and FD users by up to $50\%$ compared to the fully distributed {\QCSMA} algorithm. This delay and fairness improvement results from the different degrees of centralization at the AP. Further, we discuss the different variants and how different degrees of centralization at the AP affect the delay performance, and show that a higher degree of centralization at the AP (e.g., {\HGMSLDLE}) can result in better fairness between the FD and HD users.

To summarize, \emph{the main contribution of this paper is the design and evaluation of a distributed scheduling algorithm for infrastructure-based heterogeneous HD-FD networks that guarantees maximum throughput}. The algorithm has a relatively good delay performance and to the best of our knowledge is the first such algorithm with rigorous performance guarantees in heterogeneous HD-FD networks.

\iffullpaper
The rest of the paper is organized as follows. We discuss related work in Section~\ref{sec:related} and introduce the network model and preliminaries in Section~\ref{sec:model}. We describe the {\GMS} algorithm and develop our {\HGMS} algorithm in Section~\ref{sec:sched-alg}. The proof of throughput optimality of {\HGMS} is presented in Section~\ref{sec:mainproof}. The delay analysis of {\HGMS} and lower bounds on the average queue length are presented in Section~\ref{sec:delay}. We then illustrate the benefits of introducing FD nodes into legacy HD networks in Section~\ref{sec:benefits}. We evaluate the performance of different scheduling algorithms via simulations in Section~\ref{sec:sim} and conclude in Section~\ref{sec:conclusion}.
\fi

%% file: tex/related.tex
There has been extensive work dedicated to physical layer FD radio/system design and implementation~\cite{duarte2012experiment, bharadia2013full, chung2015prototyping, zhou2017integrated, chen2019wideband} (see also the review in~\cite{sabharwal2014band} and references therein), and open-access FD radio design based on~\cite{zhou2014low} has been integrated with the ORBIT wireless testbed~\cite{flexicon_orbit_arxiv}. Recent research also focused on characterizing and quantifying achievable throughput improvements and rate regions of FD networks in both single-channel and multi-channel cases with realistic imperfect SIC~\cite{maravsevic2016capacity,jelenaToN-FD,li2014rate}. However, these papers consider only simple network scenarios consisting of up to two links.

Most of the existing MAC layer studies focused on \emph{homogeneous} networks~\cite{goyal2013distributed, chen2017probabilistic, sahai2011pushing, xie2014does, tang2015duplex, yang2015scheduling} considering signal-to-noise ratio (SNR) or a specific standard (e.g., IEEE 802.11). For example, \cite{chen2017probabilistic} considered an IEEE 802.11 network with an FD-capable AP and HD users, and proposed an SNR-based distributed MAC protocol. As another example,~\cite{goyal2013distributed} considered an all-FD network and proposed a distributed MAC protocol based on the 802.11 DCF. 
Most relevant to our work are~\cite{yang2015scheduling} and~\cite{alim2017band} in terms of the applied techniques and network model, respectively. In particular,~\cite{yang2015scheduling} proposed a {\QCSMA}-based throughput-optimal scheduling algorithm with FD cut-through transmission in all-FD multi-hop networks, where how different classes of users (HD and FD) are affected by FD transmissions is not studied. On the other hand,~\cite{alim2017band} proposed a MAC layer algorithm for a heterogeneous HD-FD network and analyzed its throughput based on the IEEE 802.11 distributed coordination function (DCF) model~\cite{bianchi2000performance}. To the best of our knowledge, the fairness between users that have different HD/FD capabilities was not considered before.


%% file: tex/model.tex
\subsection{Network Model}

We consider a single-channel, \emph{heterogeneous} wireless network consisting of one AP and $\numUser$ users, with a UL and a DL between each user and the AP. The set of users is denoted by $\setUser$. The AP is FD, while $\numUserFD$ of the users are FD and $\numUserHD=\numUser-\numUserFD$ are HD. Without loss of generality, we index the users by $[\numUser] = \{1,2,\cdots,\numUser\}$ where the first $\numUserFD$ indices correspond to FD users and the remaining $\numUserHD$ indices correspond to HD users. The sets of FD and HD users are denoted by $\setUserFD$ and $\setUserHD$, respectively. We consider a collocated network where the users are within the communication range of each other and the AP. The network can be represented by a directed star graph $G = (\setNode, \setLink)$ with the AP at the center and two links between AP and each user in both directions. Thus, we have $\setNode = \{\textrm{AP}\} \cup \setUser$ (with $|\setNode| = 1+\numUser$) and $|\setLink| = 2\numUser$.

\subsection{Traffic Model, Schedule, and Queues}
We assume that time is slotted and packets arrive at all UL and DL queues according to some independent stochastic process. For brevity, we will use superscript $j \in \{\textrm{u}, \textrm{d}\}$ to denote the UL and DL of a user. Let $\link_{i}^{j}$ denote link $j$ (UL or DL) of user $i$, each of which is associated with a queue $\queue_{i}^{j}$. We use $\arr_{i}^{j}(t) \leq \arr_{\textrm{max}} < \infty$ to denote the number of packets arriving at link $j$ (UL or DL) of user $i$ in slot $t$. The arrival process is assumed to have a well-defined long-term rate of $\arrRate_{i}^{j} = \lim_{T \to +\infty} \frac{1}{T} \littlesum_{t=1}^{T} \arr_{i}^{j}(t)$. Let $\arrRateVec = [\arrRateUL{i}, \arrRateDL{i}]_{i=1}^{\NumUser}$ be the arrival rate vector on the ULs and DLs.

All the links are assumed to have capacity of one packet per time slot and the SIC at all the FD-capable nodes is \emph{perfect}.\footnote{We remark that imperfect SIC can also be incorporated into the model by letting the corresponding link capacity be $c_{i}^{j} \in (0,1)$. For simplicity and analytical tractability, we assume $c_{i}^{j} = 1,\ \forall  i \in \setUser$, throughout this paper.} A \emph{schedule} at any time slot $t$ is represented by a vector 
$$
\schedVec (t)= [\schedUL{1}(t), \schedDL{1}(t), \cdots, \schedUL{\numUser}(t), \schedDL{\numUser}(t)] \in \{0,1\}^{2\numUser},
$$
where $\schedUL{i}(t)$ (resp. $\schedDL{i}(t)$) is equal to $1$ if the UL (resp. DL) of user $i$ is scheduled to transmit a packet in time slot $t$ and $\schedUL{i} = 0$ (resp. $\schedDL{i} = 0$), otherwise. We denote the set of all feasible schedules by $\setFeasSched$. Let $\baseVec{i} \in \{0,1\}^{2\numUser}$ be the $i^\mathrm{th}$ basis vector (i.e., an all-zero vector except the $i^\mathrm{th}$ element being one). Since a pair of UL and DL of the same FD user can be activated at the same time, we have:
\begin{align*}
\setFeasSched & = \left\{ \zeroVec\} \cup \{\baseVec{2i-1},\baseVec{2i}, \forall i \in \setUser\} \cup \{\baseVec{2i-1}+\baseVec{2i}, \forall i \in \setUserFD \right\}.
\end{align*}
Choosing $\schedVec(t) \in \setFeasSched$, the queue dynamics are described by:
\begin{align*}
\queue_{i}^{j}(t) & = [\queue_{i}^{j}(t-1) + \arr_{i}^{j}(t)- \sched_{i}^{j}(t)]^{+},\ \forall t \geq 1,
\end{align*}
where $[\cdot]^+=\max(0,\cdot)$. $\queueVec(t) = [\queue_{i}^{\textrm{u}}(t), \queue_{i}^{\textrm{d}}(t)]_{i=1}^{\NumUser}$ denotes the queue vector, and $\mathds{1}(\cdot)$ denotes the indicator function.

\subsection{Capacity Region and Throughput Optimality}
\label{ssec:model-cap-region}

The capacity region of the network is defined as the set of all arrival rate vectors for which there exists a scheduling algorithm that can stabilize the queues. It is known that, in general, the capacity region is the convex hull of all feasible schedules~\cite{tassiulas1992stability}. Therefore, the capacity region of the heterogeneous HD-FD network is given by $\capReg_{\textrm{HD-FD}} = \cvxHull{\setFeasSched}$, where $\cvxHull{\cdot}$ is the convex hull operator. It is easy to see that this capacity region can be equivalently characterized by the following set of linear constraints\footnote{It is straightforward to only use linear inequalities, by replacing $\max\{\arrRateUL{i}, \arrRateDL{i}\}$ with $\arrRateLink{i}$ and adding linear inequalities $\arrRateUL{i} \leq \arrRateLink{i}, \arrRateDL{i} \leq \arrRateLink{i}$.}:
\begin{align}
\label{eq:cap-reg-hybrid}
\capReg_{\textrm{HD-FD}} = \{ & \arrRateVec \in [0,1]^{|\setLink|}: \nonumber \\
\littlesum\nolimits_{i \in \setUserFD} & \max\{ \arrRateUL{i}, \arrRateDL{i} \} + \littlesum\nolimits_{i \in \setUserHD} ( \arrRateUL{i} + \arrRateDL{i} ) \leq 1 \}.
\end{align}
Let a network in which all the users and the AP are only HD-capable be the \emph{benchmark all-HD network}, whose capacity region is given by $\capReg_{\textrm{HD}} = \cvxHull{\baseVec{1}, \cdots, \baseVec{2\numUser}}$, or equivalently
\begin{align}
\capReg_{\textrm{HD}} = \{ \arrRateVec \in [0,1]^{|\setLink|}: \littlesum\nolimits_{i \in \setUser} ( \arrRateUL{i} + \arrRateDL{i} ) \leq 1 \}.
\end{align}

A scheduling algorithm is called \emph{throughput-optimal} if it can keep the network queues stable for all arrival rate vectors $\arrRateVec \in \mathrm{int}(\capReg)$, where $\mathrm{int}(\capReg)$ denotes the interior of $\capReg$.

To compare $\capReg_{\textrm{HD-FD}}$ with $\capReg_{\textrm{HD}}$ and quantify the network throughput gain when a certain number of HD users become FD-capable, similar to~\cite{maravsevic2016capacity}, we define the \emph{capacity region expansion function} $\rateImprov(\cdot)$ as follows.
Given $\arrRateVec_{0}$ on the Pareto boundary of $\capReg_{\textrm{HD}}$, the capacity region expansion function at point $\arrRateVec_{0}$, denoted by $\rateImprov(\arrRateVec_{0})$, is defined as
\begin{align}
\rateImprov(\arrRateVec_{0}) = \sup\{\zeta>0: \zeta \cdot \arrRateVec_{0} \in \capReg_{\textrm{HD-FD}}\}.
\end{align}
$\rateImprov(\cdot)$ can be interpreted as a function that scales an arrival rate vector on the Pareto boundary of $\capReg_{\textrm{HD}}$ to a vector on the Pareto boundary of $\capReg_{\textrm{HD-FD}}$, as $\numUserFD$ users become FD-capable. It is not hard to see that $\rateImprov: \capReg_{\textrm{HD}} \rightarrow [1,2]$.


%% file: tex/sched_alg.tex
In this section, we develop a hybrid scheduling algorithm tailored for heterogeneous HD-FD networks. We first use Local Pooling~\cite{dimakis2006sufficient,birand2012analyzing} to prove that {\GMS} is throughput-optimal in the considered networks, and therefore, {\MWS}~\cite{tassiulas1992stability} is unneeded. Based on that, we present the {\HGMS} algorithm -- a decentralized version of {\GMS} that leverages ideas from distributed {\QCSMA}~\cite{ghaderi2010design,ni2012q}. {\HGMS} uses information about the DL queues that is available at the AP, but does not require global information about the UL queues. We state the main result (Theorem~\ref{thm:stability}) about the throughput optimality of {\HGMS} and describe its various implementations with different levels of centralization. We later show (in Section~\ref{sec:sim}) that these variants of {\HGMS} have different delay performance.

\subsection{Centralized Greedy Maximal Scheduling ({\GMS})}
\label{ssec:sched-alg-GMS}

We first show that a (centralized) {\GMS}, as described in Algorithm~\ref{alg:GMS}, is throughput-optimal in \emph{any} collocated heterogeneous HD-FD network, independent of the values of $\numUserFD$ and $\numUserHD$. In this algorithm, a pair of FD UL and DL is always scheduled at the same time, as such a schedule yields a higher throughput than scheduling only the UL or only the DL.
\begin{algorithm}[!t]
\caption{GMS for HD-FD Networks (in slot $t$)}
\label{alg:GMS}
\small
\begin{enumerate}[label=\arabic*., leftmargin=*]
\item Initialize $\schedVec(t) = \zeroVec$.
\item Select link $\link^{\star} \in \setLink$ with the largest queue length (i.e., $\link^{\star} = \arg \max_{i \in \setUser,\ j \in \setULDL} \{\queue_{i}^{j}(t)\}$). If the longest queue is not unique, break ties uniformly at random.
\item
\begin{itemize}[leftmargin=*]
\item If $\link^{\star} = \linkUL{i}\ \textrm{or}\ \linkDL{i}$ for some $i \in \setUserFD$, set $\schedUL{i}(t) = \schedDL{i}(t) = 1$;
\item If $\link^{\star} = \link_{i}^{j}$ for some $i \in \setUserHD$ and $j \in \setULDL$, set $\sched_{i}^{j}(t) = 1$.
\end{itemize}
\item Use $\schedVec(t)$ as the transmission schedule in slot $t$.
\end{enumerate}
\end{algorithm}

\begin{proposition}
\label{prop:GMS}
The Greedy Maximal Scheduling ({\GMS}) algorithm is throughput-optimal in any collocated heterogeneous HD-FD network.
\end{proposition}

The proof (see Appendix~\ref{append:GMS}) is based on~\cite[Theorem~1]{dimakis2006sufficient}, \cite{birand2012analyzing}, and the fact that the interference graph of \emph{any} collocated heterogeneous HD-FD network satisfies the Overall Local Pooling (OLoP) conditions, which guarantee that {\GMS} is throughput-optimal.

\subsection{Hybrid-GMS ({\HGMS}) Algorithm}
\label{ssec:sched-alg-HGMS}

We now present a hybrid scheduling algorithm, {\HGMS}, which combines the concepts of {\GMS} and {\QCSMA}~\cite{ghaderi2010design,ni2012q}. Instead of approximating {\MWS}~\cite{tassiulas1992stability} in a decentralized manner (as in traditional {\QCSMA}), {\HGMS} approximates {\GMS}, which is easier to decentralize in the considered HD-FD networks. {\HGMS} leverages the existence of an AP to resolve the contention among the DL queues, since the AP has explicit information about these queues and can select one of them (e.g., the longest queue). Thus, effectively at most one DL queue needs to perform {\QCSMA} in each time slot. On the other hand, since users are unaware of the UL and DL queue states of other users and at the AP, every user needs to perform {\QCSMA} in order to share the channel distributedly. Therefore, the number of possible participants under {\HGMS} in each slot is at most $(\numUser+1)$. This hybrid approach yields much better delay performance than {\QCSMA} while still achieving throughput optimality, whose proof is vastly different than that of the pure {\QCSMA}.

\begin{algorithm}[!t]
\caption{{\HGMS} Algorithm (in slot $t$)}
\label{alg:HGMS}
\small
-- If $\schedData(t-1) = \zeroVec$:
\begin{enumerate}[label=\arabic*.]
\item In the initiation mini-slot, the AP computes $\optDL = \arg \max_{i \in \setUser} \queueDL{i}(t)$. 
If multiple DL queues have the same length, break ties according to some deterministic rule. The AP chooses an initiator link $\textrm{IL}(t)$ from $\mathcal{L}(t) = \{\linkUL{1},\cdots,\linkUL{\numUser},\linkDL{\optDL}\}$ according to an access probability distribution $\probAccVec = [\probAccLink{1},\cdots,\probAccLink{\numUser},\probAccLink{\textrm{AP}}]$.
\item
If $\textrm{IL}(t) = \linkDL{\optDL}$, the AP sets:
\begin{itemize}
\item $\schedDL{\optDL}(t) = 1$ with probability $\probTXDL{\optDL}(t)$, or $\schedDL{\optDL}(t) = 0$ with probability $\probBarTXDL{\optDL}(t) = 1-\probTXDL{\optDL}(t)$;
\item In the coordination mini-slot, AP broadcasts a control packet containing the information of $\textrm{IL}(t)$ and user $\optDL$ sets $\schedUL{\optDL}(t) = \schedDL{\optDL}(t)\cdot\mathds{1}(\optDL \in \setUserFD)$;
\end{itemize}
\item
If $\textrm{IL}(t) = \linkUL{i}$ for some $i \in \setUser$, in the coordination mini-slot, the AP broadcasts the information of $\textrm{IL}(t)$ and user $i$ sets:
\begin{itemize}
\item $\schedUL{i}(t) = 1$ with probability $\probTXUL{i}(t)$, or $\schedUL{i}(t) = 0$ with probability $\probBarTXUL{i}(t) = 1-\probTXUL{i}(t)$;
\item In the same coordination mini-slot, user $i$ sends a control packet containing this information to the AP if $i \in \setUserFD$, and AP sets $\schedDL{i}(t)=\schedUL{i}(t)$;
\end{itemize}
\item At the beginning of the data slot,
\begin{itemize}
\item AP activates DL $i$ if $\schedDL{i}(t) = 1$;
\item User $i$ activates it UL if $\schedUL{i}(t) = 1$;
\end{itemize}
\end{enumerate}

-- If $\schedData(t-1) \ne \zeroVec$, set $\rm{IL}(t)=\rm{IL}(t-1)$. Repeat Steps $2$--$4$.
\end{algorithm}

Algorithm~\ref{alg:HGMS} presents the pseudocode for {\HGMS}, which operates as follows. Each slot $t$ is divided into a short control slot and a data slot. The control slot contains only two control mini-slots (independent of the number of users, $\numUser$). We refer to the first mini-slot as the \emph{initiation mini-slot} and to the second one as the \emph{coordination mini-slot}. {\HGMS} has three steps: (1) Initiation, (2) Coordination, and (3) Data transmission, as explained below.

\noindent\textbf{(1) \underline{Initiation}}. By the end of slot $(t-1)$, the AP knows $\schedVec(t-1)$ since every packet transmission has to be sent from or received by the AP. If $\schedVec(t-1) = \zeroVec$ (i.e., idle channel), then the AP starts an initiation in slot $t$ using the initiation mini-slot as follows. First, the AP centrally finds the index of the user with the longest DL queue, i.e., $\optDL(t)= \arg \max_{i \in \setUser} \queueDL{i}(t)$. If multiple DLs have equal (largest) queue length, it breaks ties according to some deterministic rule. Then, the AP randomly selects an initiator link $\textrm{IL}(t)$ from the set $\mathcal{L}(t) = \{\linkUL{1},\cdots,\linkUL{\numUser},\linkDL{\optDL}\}$ according to an \emph{access probability} distribution $\probAccVec = [\probAccLink{1},\cdots,\probAccLink{\numUser},\probAccLink{\textrm{AP}}]$ satisfying: (i) $\probAccLink{i} > 0,\forall i \in \setUser$, and $\probAccLink{\textrm{AP}}>0$, and (ii) $\probAccLink{\textrm{AP}}=1-\sum_{i=1}^{\numUser}\probAccLink{i}$. We refer to $\probAccLink{i}$ and $\probAccLink{\textrm{AP}}$ as the access probability for user $i$ and the AP, respectively. Therefore,
\begin{align}
\textrm{IL}(t) =
\begin{cases}
\linkUL{i}, & \textrm{with probability}\ \probAccLink{i},\ \forall i \in \setUser, \\
\linkDL{\optDL}, & \textrm{with probability}\ \probAccLink{\textrm{AP}},
\end{cases}
\end{align}
i.e., $\textrm{IL}(t)$ is either a UL or the DL with the longest queue. If $\schedVec(t-1) \ne \zeroVec$, set $\textrm{IL}(t) = \textrm{IL}(t-1)$.

\noindent\textbf{(2) \underline{Coordination}}. In the coordination mini-slot, if the DL of user $\optDL$ is selected as the initiator link ($\textrm{IL}(t)=\linkDL{\optDL}$), the AP sets $\schedDL{\optDL}(t)=1$ with probability $\probTXDL{\optDL}(t)$. Otherwise, it remains silent. If the AP decides to transmit on DL $\linkDL{\optDL}$ (i.e., $\schedDL{\optDL}(t)=1$), it broadcasts a control packet containing the information of $\textrm{IL}(t)$ and user $\optDL$ sets $\schedUL{\optDL}(t)=1$ if and only if $\optDL \in \setUserFD$.

If the UL of user $i$ is selected as the initiator link ($\textrm{IL}(t)=\linkUL{i}$ for some $i \in \setUser$), the AP broadcasts the information of $\textrm{IL}(t)$ and user $i$ sets $\schedUL{i}(t)=1$ with probability $\probTXUL{i}(t)$. Otherwise, user $i$ remains silent. If user $i$ is FD-capable and decides to transmit (i.e., $\schedUL{i}(t)=1$), it sends a control packet containing this information to the AP and the AP sets $\schedDL{i}(t)=1$.\footnote{Note that this operation can be done in the same coordination mini-slot since FD user $i$ can simultaneously receive the control packet ($\textrm{IL}(t)=\linkUL{i}$) from the AP and send its control packet ($\schedUL{i}(t)=1$) back to the AP.}

The transmission probability of the link is selected depending on its queue size $\queue_{i}^{j}(t)$ at the beginning of slot $t$. Specifically, similar to~\cite{ghaderi2010design,ni2012q}, link $\link_{i}^{j}$ chooses logistic form 
\begin{align}
\label{eq:tx-prob}
\probTX_{i}^{j}(t) & = \frac{\exp{(f(Q_{i}^{j}(t)))}}{1+\exp{(f(Q_{i}^{j}(t)))}},\ \forall i \in \setUser,\ \forall j \in \setULDL,
\end{align}
where $f(\cdot)$ is a positive increasing function (to be specified later), called the \emph{weight function}. 
Further, if an FD initiator UL (or DL) decides to stop transmitting (after packet transmission in the last slot), it again sends a short coordination message which stops further packet transmissions at the DL (or UL) or the same FD user.          

\noindent\textbf{(3) \underline{Data transmission}}. After steps (1)--(2), if either a pair of FD UL and DL or an HD link (UL or DL) is activated, a packet is sent on the links in the data slot. The initiator link then starts a new coordination in the subsequent control slot which either leads to more packet transmissions or stops further packet transmissions at the links involved in the schedule.

\begin{remark}
The initiation step in {\HGMS} is described as a polling mechanism where the AP draws a link $\textrm{IL}(t)$ from $\mathcal{L}(t)$ according to the access probability distribution $\probAccVec$. Alternatively, the initiation step can be described in a distributed fashion using an extra mini-slot as follows: user $i$ sends a short initiation message with probability $\probAccLink{i}$. If AP receives the message, it sends back a clear-to-initiate message and sets $\textrm{IL}(t)=\linkUL{i}$, otherwise (i.e., in case of collision or idleness) $\linkDL{\optDL}$ is selected as the initiator link by the AP. This effectively emulates polling user $i$ with probability $\widetilde{\probAcc}_{i}=\probAccLink{i}\prod_{i' \neq i}(1-\probAccLink{i'})$ and AP with probability $\widetilde{\probAcc}_{\textrm{AP}}=1-\littlesum_{i=1}^{\numUser} \widetilde{\probAcc}_{i}$.
\end{remark}

\subsection{Main Result: Throughput Optimality of {\HGMS}}
\label{ssec:sched-alg-main-result}

The system state under {\HGMS} evolves as a Markov chain $(\schedVec(t),\queueVec(t))$.  The following theorem states our main result regarding the positive recurrence of this Markov chain (throughput optimality of {\HGMS}).

\begin{theorem}
\label{thm:stability}
For any arrival rate vector $\arrRateVec \in \mathrm{int}(\capReg_{\textrm{HD-FD}})$, the system Markov chain $(\schedVec(t), \queueVec(t))$ is positive recurrent under {\HGMS} (Algorithm~\ref{alg:HGMS}). The weight function $f(\cdot)$ in {\eqref{eq:tx-prob}} can be any nonnegative increasing function such that $\lim_{x \to \infty} f(x)/\log{x}<1$, or $\lim_{x \to \infty} f(x)/\log{x}> 1$ (including $f(x)=x^{\beta},\ \beta>0$).
\end{theorem}

Establishing Theorem~\ref{thm:stability} is not trivial due to the coupling between $\schedVec(t)$ and $\queueVec(t)$: The dynamics of the schedule process $\schedVec(t)$ is governed by the queue process $\queueVec(t)$, while at the same time, the dynamics of $\queueVec(t)$ depends on $\schedVec(t)$. Depending on the functional shape of the weight function $f(\cdot)$, this coupling gives rise to vastly different behaviors for the Markov chain $(\schedVec(t),\queueVec(t))$. For functions $f(\cdot)$ that grow slower than $\log{(\cdot)}$, the convergence of the schedule process $\schedVec(t)$ occurs on a much faster time-scale (``fast mixing'') compared to the time-scale of changes in the queue process $\queueVec(t)$. For more aggressive functions $f(\cdot)$, the convergence of $\schedVec(t)$ occurs on a much slower time-scale (``slow mixing'') compared to the time-scale of changes in $\queueVec(t)$. Nevertheless, Theorem~\ref{thm:stability} states that the system Markov chain is stable (positive recurrent) for a wide range of weight functions. We provide a proof of Theorem~\ref{thm:stability} in Section~\ref{sec:mainproof} based on the analysis of the fluid limits of the system under the {\HGMS} algorithm.

\subsection{Variants of the {\HGMS} Algorithm}
\label{ssec:sched-alg-variants}

In this subsection, we introduce three variants of the {\HGMS} algorithm, which differ only in Step~1 of Algorithm~\ref{alg:HGMS}.
\begin{itemize}[leftmargin=*,topsep=0pt]
\item
{\HGMSLDL} (Algorithm~\ref{alg:HGMS}): The AP selects the \emph{longest DL}.
\item
{\HGMSRDL}: The AP selects a {DL} \emph{uniformly at random}, i.e., $\optDL \sim \mathrm{Unif}(1,\cdots,\numUser)$ (in step 1 of Algorithm~\ref{alg:HGMS}).
\item {\HGMSLDLE}: 
Exactly the same as {\HGMSLDL} except for the access probability being set according to:
\begin{align*}
\widetilde{\probAcc}_{i} & \propto \max \{ \queueEstLink{i}^{\textrm{u}} / (\littlesum\nolimits_{i'=1}^{\numUser} \queueEstLink{i'}^{\textrm{u}} + \queueDL{\optDL}),\ \probAccLink{\textrm{th}} \},\ \forall i \in \setUser, \\
\widetilde{\probAcc}_{\textrm{AP}} & \propto \max \{ \queueDL{\optDL} / (\littlesum\nolimits_{i'=1}^{\numUser} \queueEstLink{i'}^{\textrm{u}} + \queueDL{\optDL}),\ \probAccLink{\textrm{th}} \},
\end{align*}
where $\queueEstLink{i}^{\textrm{u}}$ an estimate of UL queue length of user $i$. Specifically, when a user transmits on the UL, it includes its queue length in the packets and the AP updates $\queueEstLink{i}^{\textrm{u}}$ using the most recently received UL queue length from user $i$. Then, $\probAccVec=[\probAccLink{1},\cdots,\probAccLink{\numUser},\probAccLink{\textrm{AP}}]$ is obtained after normalization, i.e.,
\begin{align*}
\probAccLink{i} & = \frac{\widetilde{\probAcc}_{i}}{\littlesum_{i'=1}^{\numUser} \widetilde{\probAcc}_{i'} + \widetilde{\probAcc}_{\textrm{AP}}},\ \forall i \in \setUser,\ 
\probAccLink{\textrm{AP}} = \frac{\widetilde{\probAcc}_{\textrm{AP}}}{\littlesum_{i'=1}^{\numUser} \widetilde{\probAcc}_{i'} + \widetilde{\probAcc}_{\textrm{AP}}}.
\end{align*}
A minimum access probability $\probAccLink{\textrm{th}}>0$ has been introduced to ensure that each link is selected with a non-zero probability. Otherwise, an HD UL $\linkUL{i}$ ($\forall i \in \setUserHD$) with a zero queue-length estimate would never be selected by the AP (i.e., $\queueEstLink{i}^{\textrm{u}} = 0$ and thus $\widetilde{\probAcc}_{i} = 0$), and the AP would never receive any updated information of $\queueEstLink{i}^{\textrm{u}}$ since $\widetilde{\probAcc}_{i}$ would remain zero.

\end{itemize}

\noindent The access probability distribution $\probAccVec$ is \emph{non-adaptive} in {\HGMSLDL} and {\HGMSRDL}, and is \emph{adaptive} in {\HGMSLDLE}. As we will see in Section~\ref{sec:sim}, the adaptive choice of $\probAccVec$ helps balance the queue lengths between FD and HD users.


%% file: tex/proof_stability.tex
We prove Theorem~\ref{thm:stability} based on the analysis of the fluid limits of the system under {\HGMS} (Algorithm~\ref{alg:HGMS}). The proof has three parts: (i) existence of the fluid limits (Lemma~\ref{lem:existence}), (ii) deriving the fluid limit equations for the various choices of $f(\cdot)$ (Lemma~\ref{lem:fluid-limit-eq}), and (iii) proving the stability of the queues in the fluid limit using a Lyapunov method, which implies the stability of the original stochastic process. The analysis and derivations are similar to the fluid limits of CSMA algorithms~\cite{bouman2011backlog, ghaderi2014queue, feuillet2010random} but specialized to the considered heterogeneous HD-FD networks. The specialization allows us to prove throughput optimality for \emph{any} nonnegative increasing weight function $f(\cdot)$ satisfying the conditions in Theorem~\ref{thm:stability}.

\noindent\textbf{Part (i): Definition and Existence of Fluid Limits.}

Consider a scaled process $\queueVec^{(r)}(t)$ where $\queueVec^{(r)}(t)=\queueVec(rt)/r$. Note that the queue process $\queueVec$ is scaled in both time and space by a factor $r>0$. To avoid technical difficulties, we can simply work with a continuous process by linear interpolation among the values at integer time points. Suppose the scaled process, with $r>0$, starts from an initial state $\queueVec^{(r)}(0)$. Any (possibly random) limit $\mathbf{q}(t)$ of the scaled process $\queueVec^{(r)}(t)$ as $r \to \infty$ is called a fluid limit. 
The process $\queueVec^{(r)}(t)$ can be constructed as follows. At any time $t\geq 0$,
\begin{align}
\label{queue-dynamics}
{\queueVec}^{(r)}(t) = {\queueVec}^{(r)}(0) + {\overline{\arrVec}}^{(r)}(t) - {\overline{\mathbf{S}}}^{(r)}(t), 
\end{align}
where for any user $i \in \setUser$ with UL or DL $j \in \setULDL$,
\begin{align*}
{{\overline{\arr}}_i^j}^{(r)}(t) & = \frac{1}{r}\littlesum_{\tau=1}^{rt} {\arr_{i}^{j}}(\tau), \\
{\overline{S}_i^j}^{(r)}(t) & = \frac{1}{r}\littlesum_{\tau=1}^{rt} {\sched_{i}^{j}(\tau)} \mathds{1}(\queue_{i}^{j} (\tau)>0).
\end{align*}
Similarly, we denote by $\mathbf{a}(t)$ and $\mathbf{s}(t)$ the limits of the scaled processes ${\overline{\arrVec}}^{(r)}(t)$ and ${\overline{\mathbf{S}}}^{(r)}(t)$ as $r \to \infty$, respectively.
The following lemma shows that the scaled process converges to the fluid limit in a weak convergence sense, in the metric of uniform norm on compact time intervals. It is possible to show a stronger convergence (i.e., almost sure convergence uniformly over compact time intervals) in the case of $\lim_{x \to \infty} f(x)/\log{x}<1$; nevertheless, the weak convergence is sufficient for our proofs.
\begin{mylemma}[Existence of Fluid Limits]
\label{lem:existence}
Suppose $\queueVec^{(r)}(0) \to \mathbf{q}(0)$. Then any sequence $r$ has a subsequence such that $({\queueVec}^{(r)}(t), {\overline{\arrVec}}^{(r)}(t), {\overline{\mathbf{S}}}^{(r)}(t)) \Rightarrow (\mathbf{q}(t),\mathbf{a}(t),\mathbf{s}(t))$ along the subsequence. The sample paths $(\mathbf{q}(t),\mathbf{a}(t),\mathbf{s}(t))$ are Lipschitz continuous and thus differentiable almost everywhere with probability one.
\end{mylemma}
\begin{IEEEproof}
The proof is standard and follows from Lipschitz continuity of the scaled process, see, e.g.,~\cite{whitt1970weak}.
\end{IEEEproof}




\noindent\textbf{Part (ii): Fluid Limit Equations under {\HGMS}.}

Recall that the schedule $\schedVec(t)$ at time $t$ is determined after the Initiation and Coordination steps of Algorithm~\ref{alg:HGMS}. Let $Y(t)$ indicate the initiator link which is activated in slot $t$. Let $\optDL = \arg \max_{i \in \setUser} \queueDL{i}(t)$, then the state space of $Y(t)$ can be labeled as $S_{Y}=\{0,1,\cdots,\numUser,\optDL\}$, where  $Y(t)=0$ means no link is active, $Y(t)=\optDL$ means DL $\linkDL{\optDL}$ is active, and $Y(t)=i$, for $i \in \{1,\cdots,\numUser\}$, means UL $\linkUL{i}$ is active.
We further use $\{Y^{\queueVec}(t)\}_{t \geq t_0} $ to denote the dynamics of $Y(t)$, assuming a fixed queue length vector $\queueVec(t)=\queueVec(t_0)=\queueVec$ for all times $t \geq t_0$. Under the {\HGMS} algorithm, $\{Y^{\queueVec}(t)\}_{t \geq t_0}$ evolves as an irreducible and aperiodic Markov chain over the state space $S_Y$. 
If $Y^{\queueVec}(t)=i$ for an $i$ which is an initiator UL or DL of an FD user $i \in \setUserFD$, then the other link of the same FD user will follow the initiator link and become active as well under {\HGMS}. Due to the activation/deactivation coordination among the initiator link and the follower link, adding the possible follower link does not change the subsequent dynamics of the Markov chain $Y^{\queueVec}(t)$ under fixed $\queueVec$.

Let $\transProbMat^{\queueVec} = [\transProb{s}{s'}]$ be the transition probability matrix of $Y^{\queueVec}(t)$, where $\transProb{s}{s'}$ is the transition probability from state $s \in S_{Y}$ to $s' \in S_{Y}$. Then, under Algorithm~\ref{alg:HGMS}, we have
\begin{equation}
\label{eq:DTMC-trans-prob}
\begin{gathered}
\transProb{0}{i} = \probAccLink{i}\probTXUL{i},\ \transProb{i}{i} = \probTXUL{i},\ ~\transProb{i}{0} = \probBarTXUL{i},\ \forall i \in \setUser \\
\transProb{0}{\optDL} = \probAccLink{\textrm{AP}}\probTXDL{\optDL},\ \transProb{\optDL}{\optDL} = \probTXDL{\optDL},\ \transProb{\optDL}{0} = \probBarTXDL{\optDL},\\
\transProb{0}{0} = 1 - \littlesum\nolimits_{i=1}^{\numUser}\transProb{0}{i} - \transProb{0}{\optDL}.
\end{gathered}
\end{equation}
\begin{mylemma}
\label{lem:DTMC-steady-state}
The steady-state distribution of Markov chain $Y^{\queueVec}(t)$ is given by
\begin{equation}
\label{eq:DTMC-steady-state}
\begin{gathered}
\pi^{\queueVec}(i) = \probAccLink{i} \exp({f(\queueUL{i})}) / Z,\ i \in S_{Y} \setminus \{0, \optDL\}; \\
\pi^{\queueVec}(\optDL) = \probAccLink{\textrm{AP}} \exp({f(\queueDL{\optDL})}) / Z,\
\pi^{\queueVec}(0) = 1/Z,
\end{gathered}
\end{equation}
where $Z$ is the normalizing constant and $f(\cdot)$ is the weight function from (\ref{eq:tx-prob}).
\end{mylemma}
\iffullpaper
The proof of Lemma~\ref{lem:DTMC-steady-state} is in Appendix~\ref{append:DTMC-steady-state}.
\fi
The following corollary is immediate as the result of Lemma~\ref{lem:DTMC-steady-state} and the fact that $Y(t)$ uniquely determines $\schedVec(t)$ by (possible) activation of both the UL and DL of an FD user in the coordination step.
\begin{corollary}
\label{cor:DTMC-steady-state}
Let $\schedSymFDBD{i} = \baseVec{2i-1}+\baseVec{2i}$, $i \in \setUserFD$, be an FD bi-directional transmission schedule, and $\schedSymHDUL{i} = \baseVec{2i-1}$ ($\schedSymHDDL{i} = \baseVec{2i}$), $i \in \setUserHD$, be an HD UL (DL) transmission schedule. Given a fixed queue vector $\queueVec(t)=\queueVec$, in steady state, if $\optDL \in \setUserFD$,
\begin{align*}
&\myProb{\schedVec = \schedSymFDBD{\optDL}}  = [ \probAccLink{\textrm{AP}}\exp(f(\queueDL{\optDL})) + \probAccLink{\optDL}\exp(f(\queueUL{\optDL})) ] / Z, \nonumber \\
&\myProb{\schedVec = \schedSymFDBD{i}}  = \probAccLink{i}\exp(f(\queueUL{i})) / Z, ~\forall i \in \setUserFD, ~i \ne \optDL, \nonumber \\
&\myProb{\schedVec = \schedSymHDUL{i}}  = \probAccLink{i}\exp(f(\queueUL{i})) / Z, ~\forall i \in \setUserHD. \nonumber
\end{align*}
Otherwise, if $\optDL \in \setUserHD$,
\begin{align*}
&\myProb{\schedVec = \schedSymFDBD{i}}  = \probAccLink{i}\exp(f(\queueUL{i})) / Z, ~\forall i \in \setUserFD, \nonumber \\
&\myProb{\schedVec = \schedSymHDUL{i}}  = \probAccLink{i}\exp(f(\queueUL{i})) / Z, ~\forall i \in \setUserHD, \nonumber \\
&\myProb{\schedVec = \schedSymHDDL{\optDL}}  = \probAccLink{\textrm{AP}}\exp(f(\queueDL{\optDL})) / Z, \notag
\end{align*}
where $Z$ and $f(\cdot)$ are as in Lemma~\ref{lem:DTMC-steady-state}.
\end{corollary}

Consider a fluid sample path under our {\HGMS} algorithm. Suppose $\mathbf{q}(t)=\mathbf{q} \neq \mathbf{0}$ at a time $t$. 
This implies that for $r$ large enough, all the queues with non-zero fluid limit $q_{i}^{j}>0$ are of size $Q_{i}^{j}= \mathcal{O}(q_{i}^{j} r)$ in the original process, while all the queues with zero fluid limit are of size $Q_{i}^{j}=o(r)$ in the original process. Therefore, taking the limit $r \to \infty$ in {\eqref{eq:DTMC-steady-state}}, and noting that the weight function $f(\cdot)$ is a positive increasing function of the queue size, it follows that
\begin{align*}
\pi^{\queueVec}(i) & \to 0\ \mbox{if } q_{i}^{\textrm{u}}=0 ,\ i \in S_{Y} \setminus \{0, \optDL\}, \\
\pi^{\queueVec}(\optDL) & \to 0\ \mbox{if } q_{\optDL}^{\rm d} = 0,\ \pi^{\queueVec}(0)=0.
\end{align*}
This shows that a queue with a zero fluid limit \emph{cannot} initiate transmission in steady state. Consequently,
\begin{align*}
\myProb{\schedVec = \schedSymFDBD{i}}  &\to 0,\ \mbox{if } \max\{q_{i}^{\textrm{u}}, q_{i}^{\textrm{d}}\}=0,\ i \in \setUserFD, \nonumber \\
\myProb{\schedVec = \schedSymHDUL{i}}  &\to 0,\ \mbox{if } q_i^{\rm u}=0, \ i \in \setUserHD, \nonumber \\
\myProb{\schedVec = \schedSymHDDL{i}}  &\to 0, \ \mbox{if } q_i^{\rm d}=0,\ i \in \setUserHD.
\end{align*}
Hence, in steady state, with high probability, the Markov chain $\schedVec(t)$ never activates an HD link with empty fluid limit queue or an FD link whose both UL and DL queues are empty, i.e., it chooses a \textit{Maximal Schedule} over the non-zero fluid queues (note that the returned schedule might not be a {\MWS} schedule). However, as mentioned in Section~\ref{ssec:sched-alg-main-result}, the Markov chain $\schedVec(t)$ might not always be at its steady state due to coupling between $\schedVec(t)$ and $\queueVec(t)$.
This coupling gives rise to qualitatively different fluid limits, depending on the time-scale of convergence of the schedule process compared to the time-scale of the changes in the queue process. For weight functions $f(\cdot)$, such that  $\lim_{r\to\infty} f(r)/\log{r}<1$, the schedule process $\schedVec(t)$ is always close to its steady state at the fluid scale; while for functions $f(\cdot)$ with $\lim_{r\to\infty} f(r)/\log{r}>1$, this does not happen. Nevertheless, in both cases, the following Lemma establishes a set of equations that the fluid limit sample paths under {\HGMS} algorithm must satisfy. The equations do not uniquely describe the fluid limit process but are sufficient to establish stability in our setting.

\begin{mylemma}[Fluid Limit Equations]
\label{lem:fluid-limit-eq}
Consider any nonnegative increasing weight function $f(\cdot)$ in {\eqref{eq:tx-prob}}, such that $\lim_{x\to\infty} f(x)/\log{x}<1$, or $\lim_{x\to\infty} f(x)/\log{x}>1$ (including $f(x)=x^{\beta},\ \beta>0$). Let  $\widehat{q}_{i}(t)=\max\{q_{i}^{\textrm{u}}(t),q_{i}^{\textrm{d}}(t)\}$, for $i \in \setUserFD$. At any regular point $t$ (i.e., any point where the derivatives of all the functions exist), for any $j \in \setULDL$,
\begin{align}
& q_{i}^{j}(t) = q_{i}^{j}(0) + a_{i}^{j}(t) - s_{i}^{j}(t),\ i \in \setUser \label{fluid1} \\
& a_{i}^{j}(t) = \lambda_{i}^{j} t,\ s_{i}^{j}(t) = \littleint\nolimits_{0}^{t} \mu_{i}^{j}(\tau)\ \textrm{d}\tau,\ \mu_{i}^{j}(t) \in [0,1], \label{fluid2} \\
&\mu_{i}^{j}(t) \cdot \mathds{1}(q_{i}^{j}(t)=0, \mathbf{q}(t) \neq \mathbf{0}) = 0,\ i \in \setUserHD, \label{fluid4} \\
&\mu_{i}^{j}(t) \cdot \mathds{1}(\widehat{q}_{i}(t)=0, \mathbf{q}(t) \neq \mathbf{0}) = 0,\ i \in \setUserFD, \label{fluid5} \\
& \textrm{if}\ q_{i}^{j}(t) = \widehat{q}_{i}(t),\ \mu_{i}^{j}(t) =\max \{\mu_{i}^{\textrm{u}}(t),\mu_{i}^{\textrm{d}}(t)\},\ i \in \setUserFD, \label{fluid5b} \\
& \textrm{if}\ \mathbf{q}(t) \neq \mathbf{0},\ \textrm{then} \nonumber \\
&\littlesum_{i\in \setUserFD} \max\{\mu_{i}^{\textrm{u}}(t), \mu_{i}^{\textrm{d}}(t)\}+ \littlesum_{i \in \setUserHD} (\mu_{i}^{\textrm{u}}(t)+\mu_{i}^{\textrm{d}}(t)) = 1. \label{fluid6}
\end{align}
\end{mylemma}
\iffullpaper
The proof of Lemma~\ref{lem:fluid-limit-eq} is provided Appendix~\ref{append:fluid-limit-eq}. 
Essentially, {\eqref{fluid1}}--{\eqref{fluid2}} hold for any scheduling algorithm and their proof is standard. $\mu_i^j(t)$ is the rate that queue $q_i^j(t)$ is served at time $t$ in the fluid limit.   
{\eqref{fluid4}}--{\eqref{fluid6}} imply that {\HGMS} chooses a maximal schedule from the non-zero fluid queues at any time, however the choice of maximal schedule could be random over the space of such maximal schedules at any time.   
\fi

\noindent\textbf{Part (iii): Stability of the Queues in the Fluid Limit.}

The following proposition proves the stability of the queues in the fluid limit, which completes the proof of Theorem~\ref{thm:stability}.

\begin{proposition}\label{lem:queue-stability}
Starting from an initial queue size $\mathbf{q}(0)$, there is a deterministic finite time $T$ by which all the queues at the fluid limit will reach zero.
\end{proposition}
\begin{IEEEproof}
Let $\widehat{q}_i(t)=\max\{q_i^{\textrm{u}}(t),q_i^{\textrm{d}}(t)\}$, $i \in \setUserFD$. Consider the Lyapunov function
\begin{align*}
V(\mathbf{q}(t)) = \littlesum\nolimits_{i \in \setUserFD} \widehat{q}_i(t) + \littlesum\nolimits_{i \in \setUserHD}(q_{i}^{\textrm{u}}(t)+ q_{i}^{\textrm{d}}(t)).
\end{align*}
Let $\mathcal{U}^j_H(t) := \{i\in \setUserHD: q_{i}^{j}(t)>0\}$, $j \in \setULDL$, and $\mathcal{U}_F(t) := \{i\in \setUserFD: \widehat{q}_i(t)>0\}$. Suppose $V(\mathbf{q}(t))>0$ (i.e., $\mathbf{q}(t) \neq \zeroVec$). Then based on the fluid limit equations {\eqref{fluid4}}--{\eqref{fluid6}}:
\begin{enumerate}
\item[(i)] The network is draining some subsets $\mathcal{P}_{H}^{\textrm{u}}(t) \subseteq \mathcal{U}_{H}^{\textrm{u}}(t)$, $\mathcal{P}_{H}^{\textrm{d}}(t) \subseteq \mathcal{U}_{H}^{\textrm{d}}(t)$, and $\mathcal{P}_{F}(t) \subseteq \mathcal{U}_{F}(t)$ of non-zero queues,
\item[(ii)] $\widehat{q}_{i}(t)$ for user $i \in \mathcal{P}_{F}(t)$ is always drained at rate $\max\{\mu_{i}^{\textrm{u}}(t),\mu_{i}^{\textrm{d}}(t)\}$,
\item[(iii)] $\littlesum_{i\in \mathcal{P}_F(t)} \max\{\mu_{i}^{\textrm{u}}(t), \mu_{i}^{\textrm{d}}(t)\}+ \littlesum_{i\in \mathcal{P}_H^{\textrm{u}}(t)} \mu_{i}^{\textrm{u}}(t)+ \littlesum_{i\in \mathcal{P}_H^{\textrm{d}}(t)} \mu_{i}^{\textrm{d}}(t)=1.$
\end{enumerate}
Hence, using {\eqref{fluid1}}--{\eqref{fluid2}} and properties (i)--(iii) above,
\begin{align*}
& \textrm{d} V(\mathbf{q}(t))/\textrm{d} t \leq \littlesum\nolimits_{i \in \setUserFD} \max\{\arrRateUL{i}, \arrRateDL{i}\} + \littlesum\nolimits_{i \in \setUserHD}(\arrRateUL{i} + \arrRateDL{i}) \\
& \quad - \littlesum\nolimits_{i\in \mathcal{P}_F(t)} \max\{\mu_{i}^{\textrm{u}}(t), \mu_{i}^{\textrm{d}}(t)\}-\littlesum\nolimits_{i\in \mathcal{P}_H^{\textrm{u}}(t)} \mu_{i}^{\textrm{u}}(t) \\
& \quad - \littlesum\nolimits_{i\in \mathcal{P}_H^{\textrm{d}}(t)} \mu_{i}^{\textrm{d}}(t)\\
& = \littlesum\nolimits_{i \in \setUserFD} \max\{\arrRateUL{i}, \arrRateDL{i}\} + \littlesum\nolimits_{i \in \setUserHD}(\arrRateUL{i} + \arrRateDL{i}) - 1 \leq -\delta,
\end{align*}
where the last inequality is due to the fact that $\arrRateVec \in \mathrm{int}(\capReg_{\textrm{HD-FD}})$, by the assumption of Theorem~\ref{thm:stability}. Thus, there must exist a small $\delta>0$ such that $\arrRateVec/(1-\delta) \in \capReg_{\textrm{HD-FD}}$. Therefore, $V(\mathbf{q}(t))$ will hit zero in finite time $T = V(\mathbf{q}(0))/\delta$, and in fact remains zero afterwards.
\end{IEEEproof}
Proposition~\ref{lem:queue-stability} implies the stability (positive recurrence) of the original Markov chain $(\schedVec(t),\queueVec(t))$ in a similar fashion as~\cite{dai1995positive} (note that the component $\schedVec(t)$ lives in a finite state space). This completes the proof of Theorem~\ref{thm:stability}.

\begin{remark}
We emphasize that, unlike the classical {\QCSMA} that approximates {\MWS} in a distributed manner, the proposed {\HGMS} algorithm approximates {\GMS} in a distributed manner. Further, we are able to establish throughput optimality for (almost) any increasing weight function $f(\cdot)$. 
\end{remark}


%% file: tex/delay.tex
In this section, we analyze the delay performance of {\HGMS} in terms of the average queue length in order to provide a benchmark for the performance evaluation in Section~\ref{sec:sim}. In particular, we derive two lower bounds: (i) a fundamental lower bound that is independent of the scheduling algorithms, and (ii) an improved lower bound tailored for the developed {\HGMSLDL} and {\HGMSRDL}.\footnote{The analysis can possibly be extended to {\HGMSLDLE} by incorporating its \emph{time-varying and queue-dependent} access probability. We leave this analysis for future work.} In Section~\ref{ssec:sim-delay}, we numerically evaluate these lower bounds and compare them to the average queue length achieved by various scheduling algorithms.

We adopt the following notation. Given a set of links $\mathcal{L}$, we use $\arrRateSum{\mathcal{L}} = \sum_{\link \in \mathcal{L}} \arrRateLink{\link}$ to denote the sum of arrival rates, and use $\queueSum_{\mathcal{L}} = \littlesum_{\link \in \mathcal{L}} \myE{\queueLink{\link}}$ to denote the expected sum of queue lengths of $\mathcal{L}$ in steady state. 
The average queue length in a given heterogeneous HD-FD network, $(\setUser, \setLink)$, is defined by
\begin{align}
\label{eq:avg-queue-length}
\queueAvg & = \littlesum\nolimits_{\link \in \setLink} \myE{\queueLink{\link}}/|\setLink| = \queueSum_{\setLink}/(2\numUser).
\end{align}
Therefore, finding a lower bound on $\queueAvg$ is equivalent to finding a lower bound on $\queueSum_{\setLink}$.
\subsection{A Fundamental Lower Bound}
We first derive a fundamental lower bound on $\queueAvg$ that is independent of {the chosen (possibly centralized)} scheduling algorithm, based on the following result.
\begin{proposition}[{\hspace{1sp}\cite[Proposition~4.1]{gupta2010delay}}]
\label{prop:lb-gupta-shroff}
With independent packet arrivals, the expected sum of queue lengths in a clique $\mathcal{C}$ under any scheduling policy satisfies
\begin{align*}
\queueSum_{\mathcal{C}} = \littlesum_{l \in \mathcal{C}} \myE{Q_{l}} \geq \littlesum_{l \in \mathcal{C}} \frac{\arrRateLink{l} + \myVar{\arr_{l}} - \arrRateLink{l}\arrRateSum{\mathcal{C}}}{2 \left(1-\arrRateSum{\mathcal{C}}\right)} := \queueSum_{\mathcal{C}}^{\textrm{LB}}.
\end{align*}
\end{proposition}

Note that $\queueSum_{\mathcal{C}}^{\textrm{LB}}$ is equivalent to the sum of queue lengths in a standard single-server GI/D/1 queue in clique $\mathcal{C}$. In order to obtain a tight fundamental lower bound in the heterogeneous HD-FD networks, one needs to find the largest clique of links, $\setLinkMax$, with the maximal sum of arrival rates. In particular, we divide $\setLink$ into two disjoint sets $\setLink = \setLinkMax \cup \setLinkMin$:
\begin{equation*}
\label{eq:disjoint-sets}
\begin{cases}
\setLinkMax = \{ \link_{i}^{j}: \forall i \in \setUserFD\ \textrm{if}\ \arrRateLink{i}^{j} \geq \arrRateLink{i}^{\overline{j}} \} \cup \{ \linkUL{i}, \linkDL{i}: \forall i \in \setUserHD \}, \\
\setLinkMin = \{ \link_{i}^{j}: \forall i \in \setUserFD\ \textrm{if}\ \arrRateLink{i}^{j} < \arrRateLink{i}^{\overline{j}} \},
\end{cases}
\end{equation*}
where {$\{\overline{j}\} = \setULDL \setminus \{j\}$} and we break ties uniformly at random if $\arrRateUL{i} = \arrRateDL{i}$ for $\forall i \in \setUserFD$. {Essentially, $\setLinkMax$ includes the UL \emph{and} DL of each HD user, and the higher arrival rate link (UL \emph{or} DL) of each FD user. As a result, $\arrRateSum{\setLinkMax}$ approaches $1$ as $\arrRateVec$ approaches the boundary of $\capReg_{\textrm{HD-FD}}$ (see {\eqref{eq:cap-reg-hybrid}}). The following proposition gives the fundamental lower bound on the average queue length in the heterogeneous HD-FD networks.}

\begin{proposition}
\label{prop:delay-lb-fundamental}
A fundamental lower bound on the average queue length in the considered heterogeneous HD-FD networks, denoted by $\queueAvg_{\textrm{Fund}}^{\textrm{LB}}$, is given by
\begin{align}
\label{eq:delay-lb-fundamental}
\queueAvg & \geq \queueAvg_{\textrm{Fund}}^{\textrm{LB}} := \queueSum_{\setLinkMax}^{\textrm{LB}}/(2\numUser),
\end{align}
{where $\queueAvg$ is the average queue length defined in {\eqref{eq:avg-queue-length}}, and $\queueSum_{\setLinkMax}^{\textrm{LB}}$ is given by Proposition~\ref{prop:lb-gupta-shroff} for clique $\setLinkMax$.}
\end{proposition}
\begin{IEEEproof}
Since a pair of FD UL and DL will always be activated at the same time, it holds that for $\forall i \in \setUserFD$, $\myE{\queue_{i}^{j}} \geq \myE{\queue_{i}^{\overline{j}}}$ if $\arrRateLink{i}^{j} \geq \arrRateLink{i}^{\overline{j}}$. By assigning the FD UL/DL with a higher arrival rate to $\setLinkMax$, we construct a maximal clique, $\setLinkMax$, with the maximal possible sum or arrival rates. Although it is possible that two queues from both $\setLinkMin$ and $\setLinkMax$ are served simultaneously, it is still guaranteed that
\begin{align}
\queueSum_{\setLink} & \geq \queueSum_{\setLinkMax} \geq \queueSum_{\setLinkMax}^{\textrm{LB}},
\end{align}
and Proposition~\ref{prop:delay-lb-fundamental} follows directly.
\end{IEEEproof}

\subsection{An Improved Lower Bound under {\HGMSLDL} and {\HGMSRDL}}
{We now derive an improved lower bound on $\queueAvg$ for the considered heterogeneous HD-FD networks taking into account the characteristics of the developed {\HGMSLDL} and {\HGMSRDL} (e.g., the access probability $\probAccVec$ and the transmission probability $\probTX(\cdot)$). The result is stated in the following proposition.}
\begin{proposition}
\label{prop:delay-lb-hgms}
{
Let $\myInv{\probTX}(\cdot)$ be the inverse of the transmission probability $\probTX(\cdot)$ given by {\eqref{eq:tx-prob}}.
Let $\arrRateLink{\textrm{min}} = \min_{i \in \setUser} \{\arrRateUL{i},\arrRateDL{i}\}$ be the minimum link arrival rate and  $\probAccLink{\textrm{max}} = \max \{\probAccLink{1},\cdots,\probAccLink{\numUser},\probAccLink{\textrm{AP}}\}$ be the maximum access probability. The average queue length under {\HGMSLDL} and {\HGMSRDL} is lower bounded by $\queueAvg_{\HGMS}^{\textrm{LB}}$ given by}
\begin{align}
\label{eq:delay-lb-hgms}
& \queueAvg \geq \queueAvg_{\HGMS}^{\textrm{LB}} := \max \Big\{ \queueAvg_{\textrm{Fund}}^{\textrm{LB}}, \nonumber \\
& \qquad\quad \Big(1-\frac{\numUserFD}{2\numUser}\Big) \cdot \probTX^{-1} \Big( \frac{ \arrRateLink{\textrm{min}}/\probAccLink{\textrm{max}}}{1 - \arrRateSum{\setLinkMax} + \arrRateLink{\textrm{min}}/\probAccLink{\textrm{max}} } \Big)
\Big\},
\end{align}
where $\queueAvg_{\textrm{Fund}}^{\textrm{LB}}$ is given in Proposition~\ref{prop:delay-lb-fundamental}.
\end{proposition}
\begin{IEEEproof}
The proof is based on the workload decomposition rules~\cite{boxma1989workloads} and can be found in Appendix~\ref{append:proof-delay-lb-hgms}.
\end{IEEEproof}
\begin{remark}
Note that \eqref{eq:delay-lb-hgms} applies to any 
variant of {\HGMS} with fixed access probability 
$\probAccVec$. The lower bound 
$\queueAvg_{\HGMS}^{\mathrm{LB}}$ depends on: (i) 
the ratio between the link arrival rate and access 
probability  
$\frac{\arrRateLink{\mathrm{min}}}{\probAccLink{\mathrm{max}}}$, and (ii) the weight function 
$f(\cdot)$ (through $p(\cdot)$). A more aggressive $f(\cdot)$ results in a 
lower value of $\queueAvg_{\HGMS}^{\mathrm{LB}}$. 
The lower bound can be also applied to {\HGMSLDLE} 
by setting $\probAccLink{\mathrm{max}} = 1$; 
however, this will result in a loose lower bound as 
it ignores the adaptive behavior of $\probAccVec$.
\end{remark}


%% file: tex/benefits.tex
In this section, we illustrate the benefits of introducing FD-capable nodes into all-HD networks, in terms of obtained throughput gains. The throughput gains can be expressed for individual users or the network (i.e., the sum rates). We define the network (individual) throughput gain as the ratio between the achievable network (individual) throughput in a heterogeneous HD-FD network and that in an all-HD network with the same total number of users.

{For simplicity and illustrative purposes, consider a static version of $\HGMSRDL$, 
with access probabilities $\probAccVec = \frac{1}{1+\numUser} \cdot \oneVec$ (see Algorithm~\ref{alg:HGMS} and Section~\ref{ssec:sched-alg-variants}), and fixed transmission probabilities $\probTXUL{f}=\probTXDL{f}=\probTXLink{f},\ \probTXUL{h}=\probTXDL{h}=\probTXLink{h} \in (0,1)$ for FD and HD users in {\eqref{eq:tx-prob}}, respectively. By analyzing the Markov chain (similar to Lemma~\ref{lem:DTMC-steady-state}) under fixed $\probAccVec$, $\probTXLink{f}$, and $\probTXLink{h}$, the network throughput (i.e., sum rates) of the heterogeneous HD-FD network, $\tput_{\textrm{HD-FD}}$, is given by}
\begin{align}
\label{eq:tput-qcsma-constant-p}
\tput_{\textrm{HD-FD}} & = \frac{\frac{2\numUserFD}{\numUser} \frac{\probTXLink{f}}{1-\probTXLink{f}} + \frac{\numUserHD}{\numUser} \frac{\probTXLink{h}}{1-\probTXLink{h}}}{1 + \frac{\numUserFD}{\numUser} \frac{\probTXLink{f}}{1-\probTXLink{f}} + \frac{\numUserHD}{\numUser} \frac{\probTXLink{h}}{1-\probTXLink{h}}}.
\end{align}

\begin{figure}[!t]
\centering
\vspace{-\baselineskip}
\subfloat{
\includegraphics[width=0.45\columnwidth]{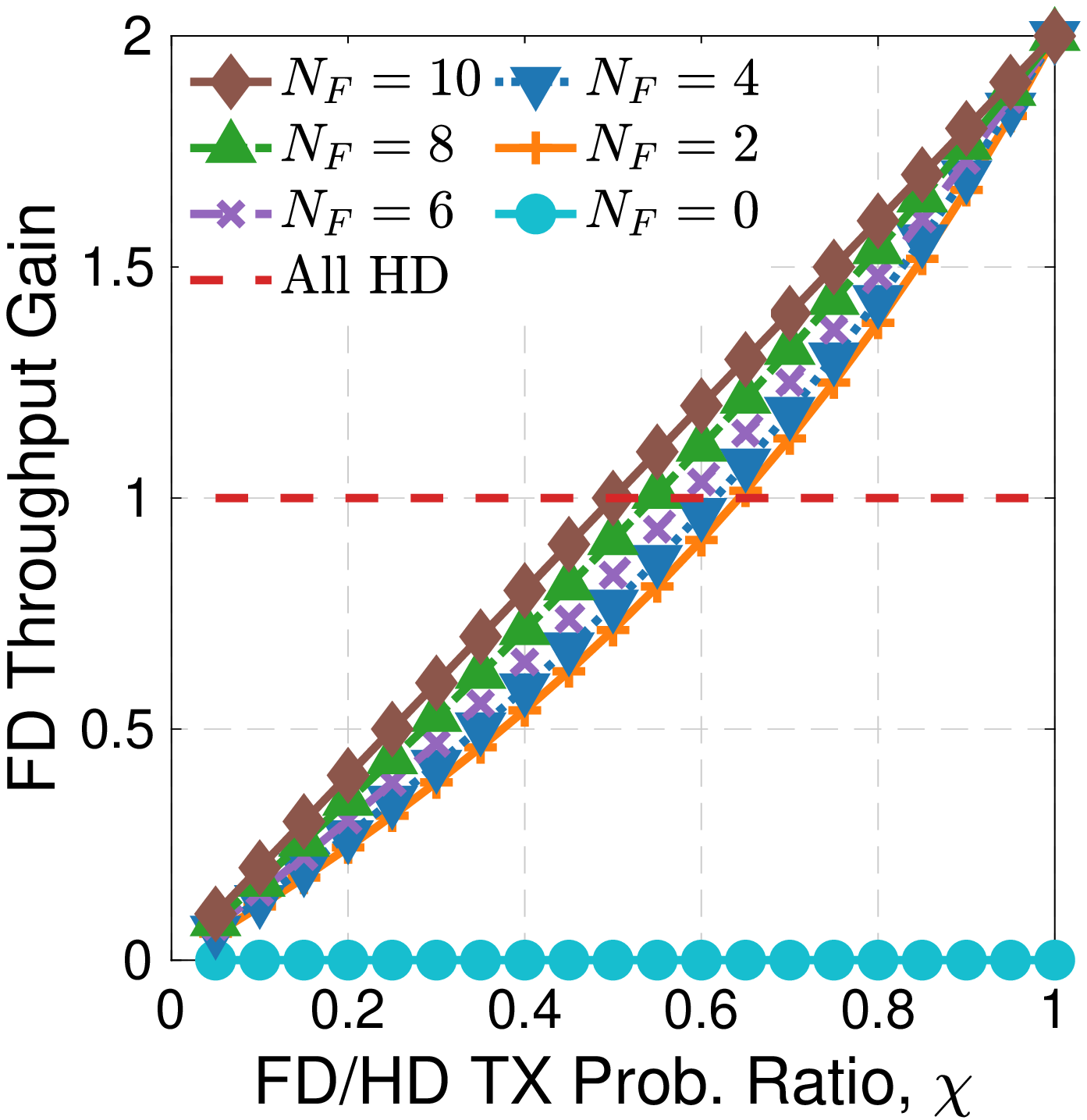}
\label{fig:qcsma-constant-p-user-gain-fd}
}
\subfloat{
\includegraphics[width=0.45\columnwidth]{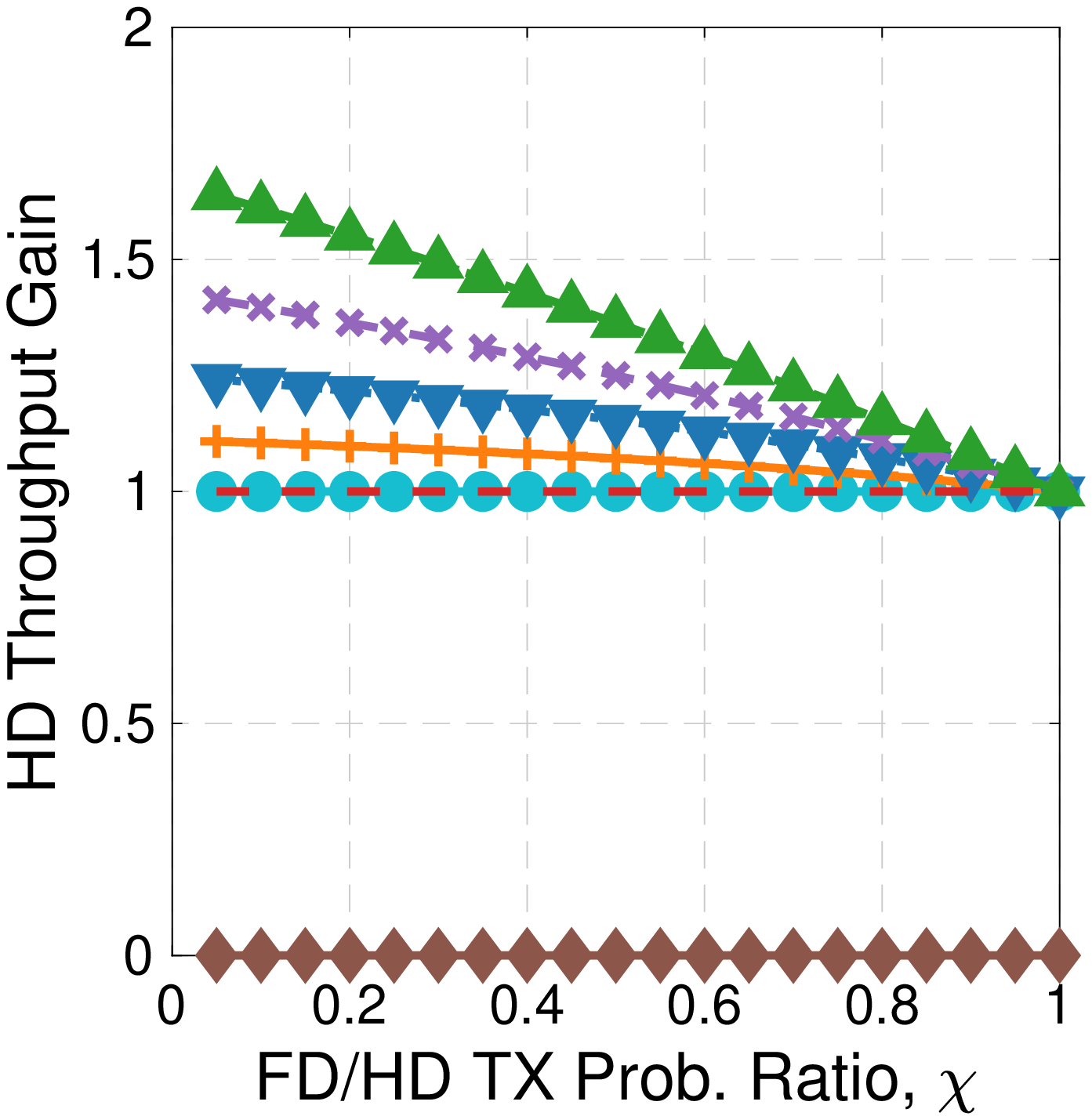}
\label{fig:qcsma-constant-p-user-gain-hd}
}
\caption{Throughput gain of FD and HD users when the throughput is compared to the individual throughput of an HD user in the all-HD network under the static {\HGMSRDL} algorithm, with $\numUser = 10$, $\numUserFD \in \{0,2,\cdots,10\}$, and $\probTXLink{h}=0.5$.}
\label{fig:qcsma-constant-p-user-gain}
\vspace{-\baselineskip}
\end{figure}

Note that the throughput of the benchmark all-HD network is simply $\tput_{\textrm{HD}} = \probTXLink{h}$. If $\probTXLink{f} = \probTXLink{h} = p$ (i.e., FD and HD users transmit with the same probability when they capture the channel), {\eqref{eq:tput-qcsma-constant-p}} becomes $S_{\textrm{HD-FD}} = (1+\frac{\numUserFD}{\numUser}) \cdot p$. This implies that under the static {\HGMSRDL}, the network throughput gain achieved by the HD-FD network is $(1+\frac{\numUserFD}{\numUser}) \in [1,2]$, which increases with respect to $\numUserFD$.

Assigning equal transmission probabilities results in FD users having $2\times$ throughput compared to the HD users. We can balance the throughput obtained by FD and HD users by assigning different transmission probabilities. Let $\probTXLink{h}=p$ and $\probTXLink{f}=\chi \cdot p$ for some \emph{transmission probability ratio} $\chi$. In order to balance the individual throughput of FD and HD users, we lower the priority of FD transmissions by choosing $\chi \in (0,1]$. 

We numerically evaluate the individual user throughput gain. We consider both the benchmark all-HD network (with transmission probability $\probTXLink{h}=p$) and HD-FD networks with $\numUser=10$ and vary $\numUserFD \in \{0,2,\cdots,10\}$ in the latter. We select constant $\probTXLink{h}=0.5$ and $\probTXLink{f}=\chi \cdot \probTXLink{h}$ with varying $\chi \in (0,1]$. Fig.~\ref{fig:qcsma-constant-p-user-gain} plots individual throughput gains of an FD or HD user. As Fig.~\ref{fig:qcsma-constant-p-user-gain} suggests, if FD and HD users are assigned equal transmission probabilities ($\chi=1$), an FD user gets $2\times$ throughput compared to an HD user. If the transmission probability of the FD users is lowered (by decreasing $\chi$), the throughput of FD and HD users is more balanced. For example, with $\chi=0.75$, the individual throughput gains of FD and HD users are $43\%$ and $20\%$, respectively.

The results reveal an interesting phenomenon: when $\numUserFD$ is sufficiently large, at the cost of slightly lowering the priority of FD users, even HD users can experience throughput improvements. This opens up the possibility of designing wireless protocols with different fairness-efficiency tradeoffs by setting different priorities among FD and HD users.


%% file: tex/simulation.tex
\begin{figure}[!t]
\centering
\includegraphics[width=\columnwidth]{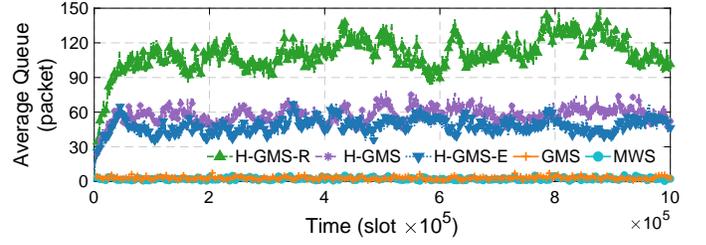}
\vspace{-\baselineskip}
\caption{Sample path of average queue length per link under different scheduling algorithms for a heterogeneous HD-FD network with $\numUserFD = \numUserHD = 5$, and \emph{very high} traffic intensity $\rho=0.95$.}
\vspace{-\baselineskip}
\label{fig:delay-sample-path}
\end{figure}

In this section, we evaluate the performance of different scheduling algorithms in heterogeneous HD-FD networks via simulations. We focus on (i) network-level \emph{delay} performance (represented by the long-term average queue length per link), and (ii) \emph{fairness} between FD and HD users (represented by the relative delay performance between FD and HD users).

\begin{figure*}[!t]
\centering
\vspace{-\baselineskip}
\subfloat[$\numUserFD=0$, $\numUserHD=10$]{
\includegraphics[width=0.6\columnwidth]{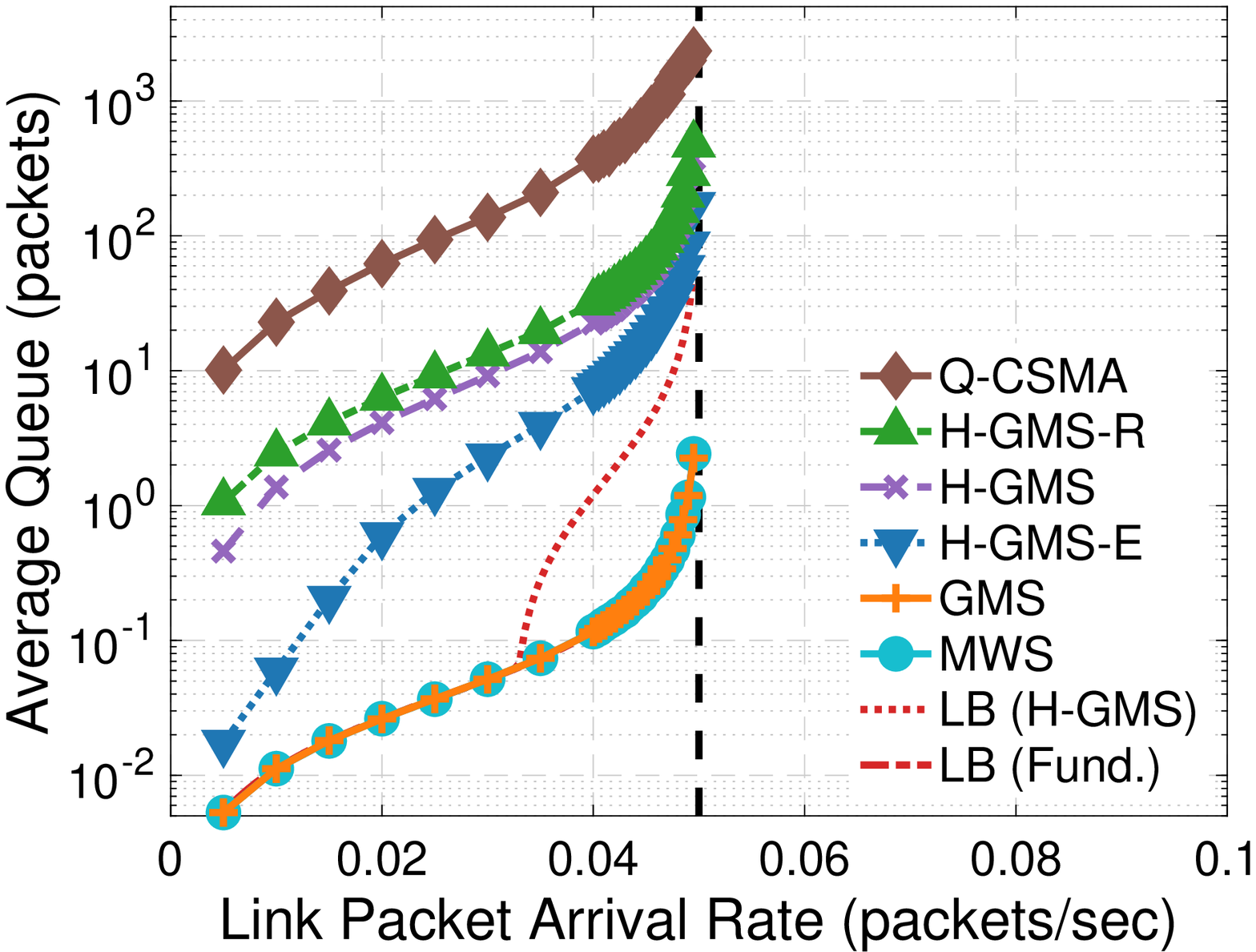}
\label{fig:delay-and-lb-n-10-nfd-0}
} \hspace{6pt}
\subfloat[$\numUserFD=5$, $\numUserHD=5$]{
\includegraphics[width=0.6\columnwidth]{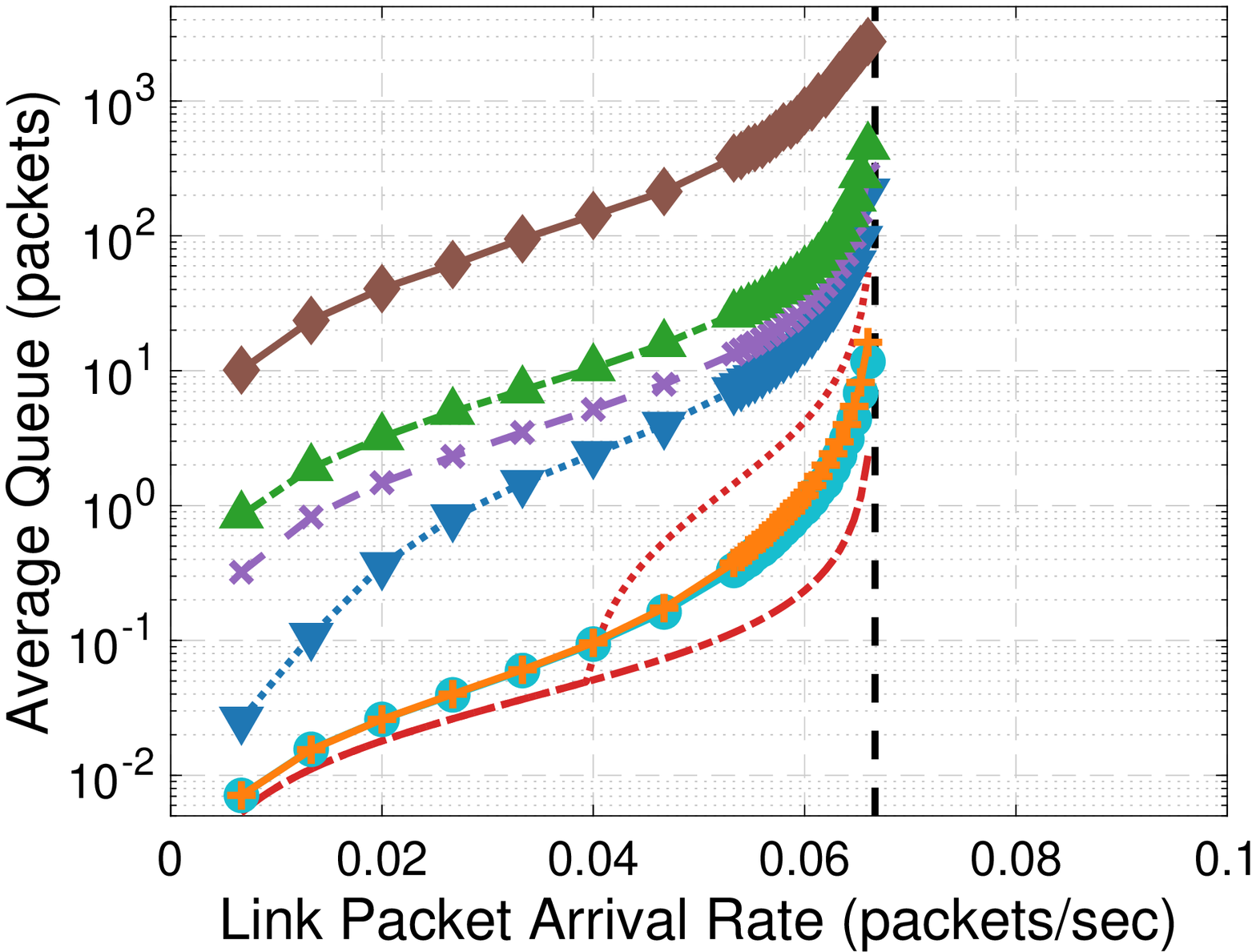}
\label{fig:delay-and-lb-n-10-nfd-5}
} \hspace{6pt}
\subfloat[$\numUserFD=10$, $\numUserHD=0$]{
\includegraphics[width=0.6\columnwidth]{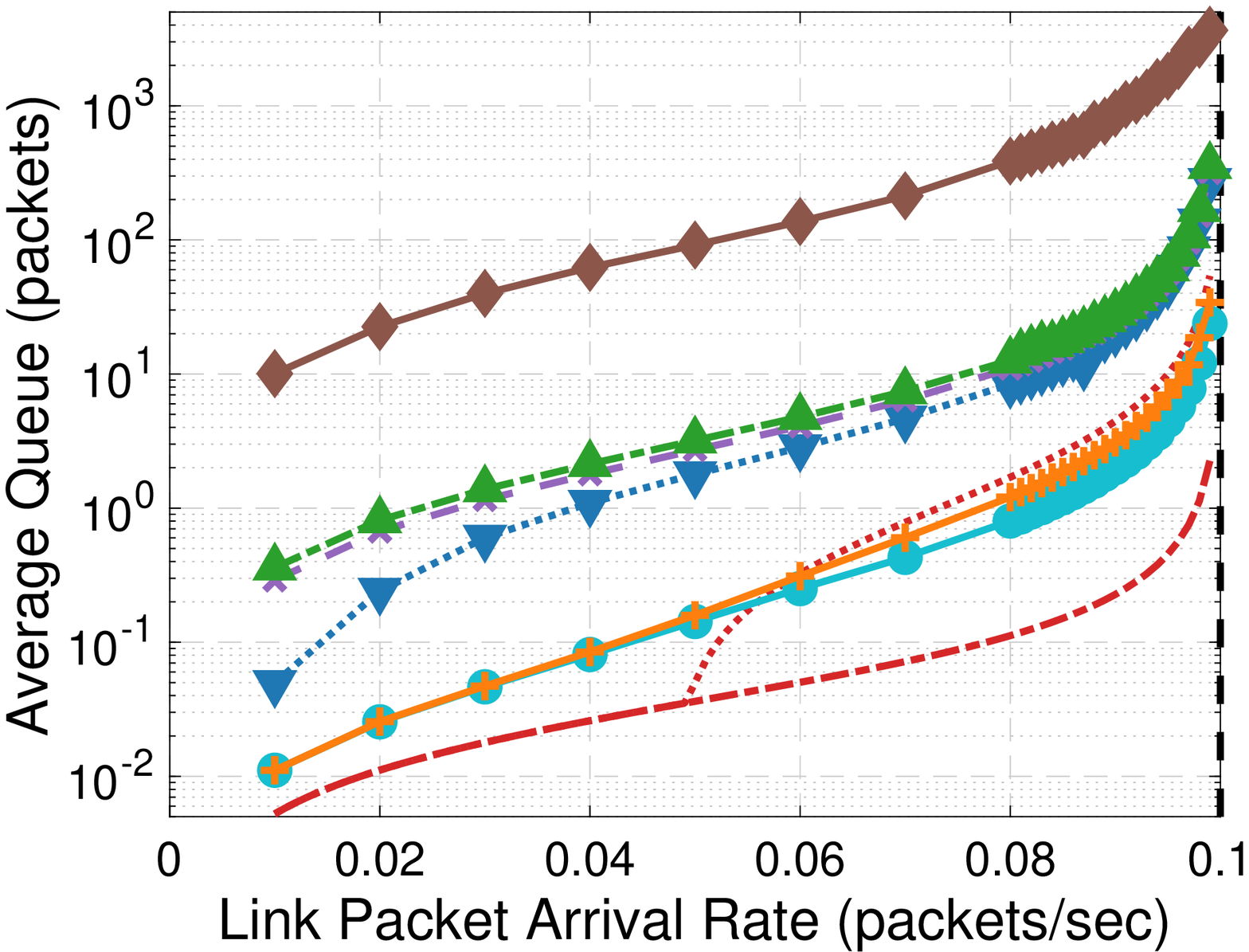}
\label{fig:delay-and-lb-n-10-nfd-10}
}
\vspace{-0.5\baselineskip}
\caption{Long-term average queue length per link in a heterogeneous HD-FD network with $\numUser = 10$ and equal arrival rates, under different scheduling algorithms and varying number of FD users, $\numUserFD$: (a) $\numUserFD = 0$, (b) $\numUserFD = 5$, and (c) $\numUserFD = 10$. Both the fundamental and improved lower bounds on the delay are also plotted according to {\eqref{eq:delay-lb-fundamental}} and {\eqref{eq:delay-lb-hgms}}. The capacity region boundary in each HD-FD network is illustrated by the vertical dashed line.}
\vspace{-1.5\baselineskip}
\label{fig:delay-and-lb}
\end{figure*}

\subsection{Setup}
\label{ssec:sim-setup}
Throughout this section, we consider heterogeneous HD-FD networks with one FD AP and $10$ users ($\numUser=10$), with a varying number of FD users, $\numUserFD$.\footnote{The results for heterogeneous HD-FD networks with a different number of users, $\numUser$, are similar, and thus, omitted.} We choose a rate vector $\mathbf{v} = [v_{i}^{\textrm{u}},v_{i}^{\textrm{d}}]_{i=1}^{N}$ on the boundary of the capacity region $\capReg_{\textrm{HD-FD}}$ (see Section~\ref{ssec:model-cap-region}) and consider arrival rates of the form $\arrRateVec = \rho \mathbf{v}$, in which $\rho \in (0,1)$ is the \emph{traffic intensity}. Note that as $\rho \to 1$, $\arrRateVec$ approaches the boundary of $\capReg_{\textrm{HD-FD}}$. Since we focus on the fairness between FD and HD users, we assume equal UL and DL arrival rates over all the users. Therefore, for $j \in \setULDL$, we use $v_{f} = v_{i}^{j}, ~\forall i \in \setUserFD$, and $v_{h} = v_{i}^{j}, ~\forall i \in \setUserHD$, to denote the equal UL and DL arrival rates assigned to FD and HD users, respectively. For an \emph{equal arrival rate} model, we have $v_{f} = v_{h} = 1/(\numUserFD+2\numUserHD)$ computed using~\eqref{eq:cap-reg-hybrid}.

The packet arrivals at each link $\link_{i}^{j}$ follow an independent Bernoulli process with rate $\arrRateLink{i}^{j}$. For each algorithm under a given traffic intensity, $\rho$, we take the average over $10$ independent simulations, each of which lasts for $10^{6}$ slots. For simplicity, we refer to the ``queue length of an FD (resp. HD) user'' as the sum of its UL and DL queue lengths, and only compare the average queue length between FD and HD users without distinguishing between individual UL and DL. The considered algorithms include:
\begin{itemize}[leftmargin=*,topsep=0pt]
\item {\MWS}, {\GMS}: The centralized MWS and GMS algorithms;
\item {\HGMSLDL}, {\HGMSRDL}, and {\HGMSLDLE}: Three variants of the {\HGMS} algorithm as described in Section~\ref{ssec:sched-alg-variants};
\item {\QCSMA}: The standard distributed {\QCSMA} algorithm from~\cite{ni2012q}, in which each link (UL or DL) performs channel contention independently and the AP does not leverage the central DL queue information.
\end{itemize}
In the last four distributed algorithms, the transmission probability of link $\link$ in slot $t$ is selected as $\probTXLink{\link}(t) = \frac{\exp{(f(\queueLink{\link}(t)))}}{1+\exp{(f(\queueLink{\link}(t)))}}$ where the weight function $f(x) = \log{(1+x)}$ (i.e., $\probTXLink{\link}(t) = \frac{1+\queueLink{\link}(t)}{2+\queueLink{\link}(t)}$). {We set $\probAccVec = \frac{1}{1+\numUser} \cdot \oneVec$ for {\HGMSLDL} and {\HGMSRDL}, and $\probAccLink{\textrm{th}} = 0.01$ for {\HGMSLDLE} (see Section~\ref{ssec:sched-alg-variants})}. We will show that different degrees of centralization at the AP result in performance improvements of {\HGMS} over the classical {\QCSMA} in terms of both delay and fairness. We also consider effects of different weight functions in Section~\ref{ssec:sim-diff-tx-prob}.

\subsection{Delay Performance}
\label{ssec:sim-delay}

We first consider the queue length dynamics under various scheduling algorithms in an HD-FD network with $\numUserFD = \numUserHD = 5$ and traffic intensity $\rho=0.95$. This implies that $v_{f} = v_{h} = 1/15$, corresponding to a capacity region expansion value of $\rateImprov = 4/3$ (see Section~\ref{ssec:model-cap-region} with $v_{h}=1/20$ in the all-HD network). Fig.~\ref{fig:delay-sample-path} plots the sample paths of the average queue length of the network (i.e., averaged over all the ULs and DLs) under different algorithms. The result for {\QCSMA} algorithm is omitted since, as we will see shortly, its average queue length is at least one order of magnitude larger than those achieved by other algorithms.

Fig.~\ref{fig:delay-and-lb} plots the average queue length with varying traffic intensities in HD-FD networks with $\numUser=10$ and $\numUserFD \in \{0,5,10\}$. Recall that in the equal arrival rate model, the relationship between the link packet arrival rate and traffic intensity is  $\arrRateLink{i}^{j} = \rho/(\numUserFD+2\numUserHD)$, $\forall i \in \setUser$, $\forall j \in \setULDL$. {Fig.~\ref{fig:delay-and-lb} shows that the capacity region of the HD-FD networks expands with increased value of $\numUserFD$.} Compared with~Figs.~\ref{fig:delay-and-lb}\subref{fig:delay-and-lb-n-10-nfd-0},  Figs.~\ref{fig:delay-and-lb}\subref{fig:delay-and-lb-n-10-nfd-5} and~\ref{fig:delay-and-lb}\subref{fig:delay-and-lb-n-10-nfd-10} show a capacity region expansion value of $\rateImprov=4/3$ for $\numUserFD=5$, and $\rateImprov=2$ for $\numUserFD=10$, respectively.

Figs.~\ref{fig:delay-sample-path} and~\ref{fig:delay-and-lb} show that, as expected from Theorem~\ref{thm:stability}, all the considered algorithms are throughput-optimal -- they stabilize all network queues. The fully-centralized {\MWS} and {\GMS} have the best delay performance but require high-complexity implementations. Among distributed algorithms, {\QCSMA}~\cite{ni2012q} has the worst delay performance due to the high contention intensity introduced by a total of $2\numUser$ contending links. By ``consolidating'' the $\numUser$ DLs into one DL that participates in channel contention, {\HGMSRDL}, {\HGMSLDL}, and {\HGMSLDLE} achieve at least $9$--$16\times$, $16$--$30\times$, and $25$--$50\times$ better delay performance than {\QCSMA}, respectively, under different traffic intensities $\rho$. In particular, {\HGMSLDL} and {\HGMSLDLE} have similar delay performance which is better than for {\HGMSRDL}, since the AP leverages its central information to always select the longest queue DL  for channel contention. However, {\HGMSLDL} and {\HGMSLDLE} provide different fairness among FD and HD users due to the choice of access probability distribution $\probAccVec$ (that is constant for the former and depends on the queue-length estimates for the latter), as we show below.

Fig.~\ref{fig:delay-and-lb} also presents both the fundamental and improved lower bounds on the delay, $\queueAvg_{\textrm{\textrm{Fund}}}^{\textrm{LB}}$ and $\queueAvg_{\textrm{\textrm{H-GMS}}}^{\textrm{LB}}$, given by {\eqref{eq:delay-lb-fundamental}} and {\eqref{eq:delay-lb-hgms}}, respectively. The turning point of $\queueAvg_{\textrm{\textrm{H-GMS}}}^{\textrm{LB}}$ where it starts to deviate from $\queueAvg_{\textrm{\textrm{Fund}}}^{\textrm{LB}}$ is because of the $\max(\cdot)$ operator in {\eqref{eq:delay-lb-hgms}}. As Fig.~\ref{fig:delay-and-lb} suggests, the fundamental lower bound, $\queueAvg_{\textrm{\textrm{Fund}}}^{\textrm{LB}}$ is very close to the average queue length obtained by {\MWS} and {\GMS} (they indeed match perfectly in the all-HD network with $\numUserHD = \numUser=10$). However, in heterogeneous HD-FD networks, $\queueAvg_{\textrm{\textrm{H-GMS}}}^{\textrm{LB}}$ provides a much tighter lower bound on the average queue length achieved by {\HGMS}, especially with high traffic intensities.

\subsection{Fairness}
\label{ssec:sim-fairness}

Our next focus is on the fairness performance of {\HGMS}. Here, we define fairness between FD and HD users as the \emph{ratio between the average queue length of FD and HD users}. We use this notion since, intuitively, if an FD user experiences lower average delay (i.e., queue length) than an HD user, then introducing FD capability to the network will imbalance the service rate both users get. Ideally, we would like the proposed algorithms to achieve good fairness performance in the considered HD-FD networks. Similarly, we define fairness between ULs and DLs as the \emph{ratio between the average UL and DL queue lengths} to evaluate the effects of different levels of centralization at the AP when operating {\HGMS}.

\subsubsection{Equal Arrival Rates}
We first evaluate the fairness under different distributed algorithms with equal arrival rates at each link. We focus on traffic intensity regime of $\rho \in [0.5,1)$ since, as shown in Fig.~\ref{fig:delay-and-lb}, all links have very small queue lengths with low traffic intensities (e.g., the average queue length is less the $10$ packets with $\rho = 0.5$).

\begin{figure}[!t]
\centering
\vspace{-0.5\baselineskip}
\subfloat[b/w FD and HD users]{
\includegraphics[width=0.45\columnwidth]{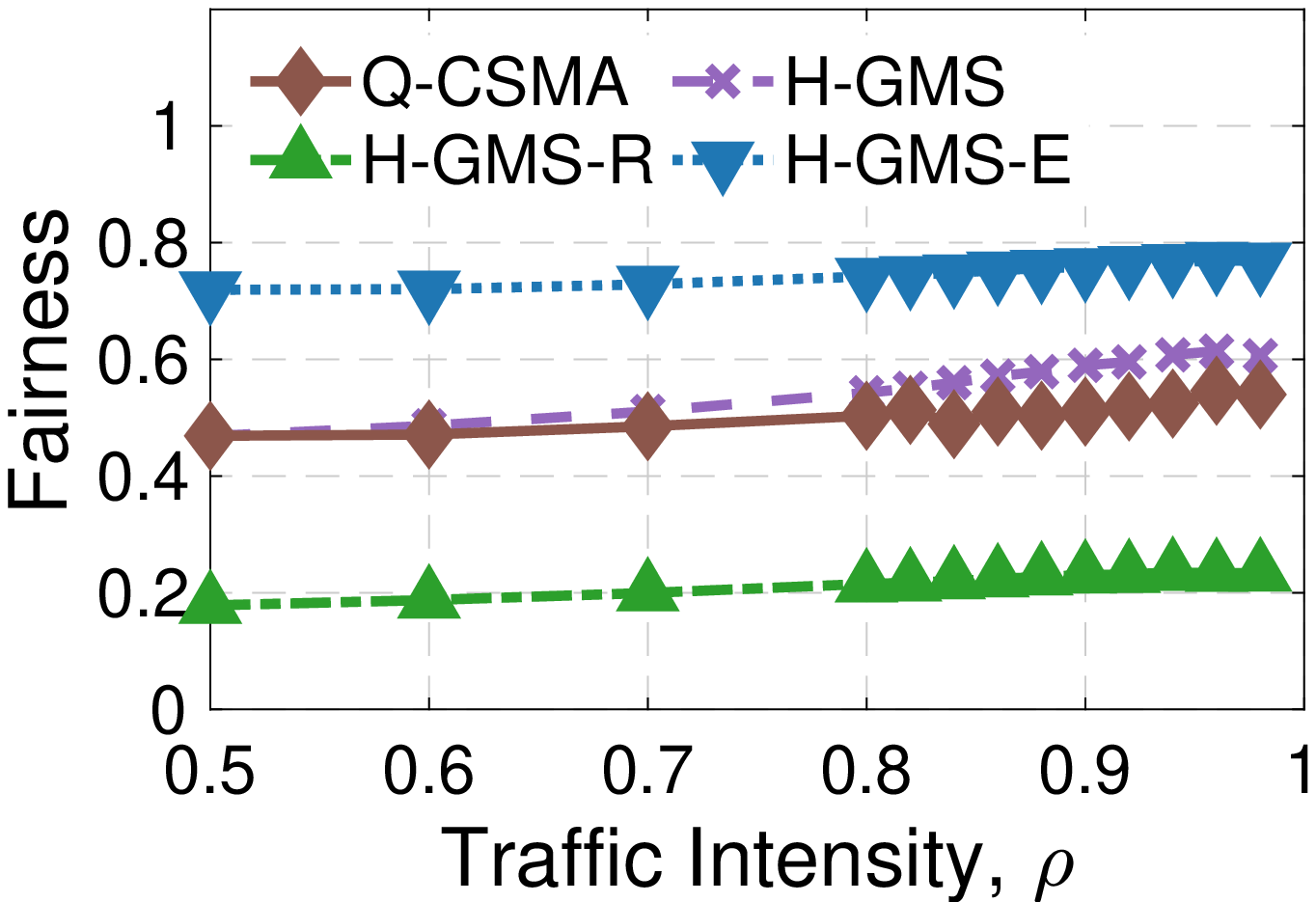}
\label{fig:fairness-equal-rate-fd-vs-hd}
}
\subfloat[b/w ULs and DLs]{
\includegraphics[width=0.45\columnwidth]{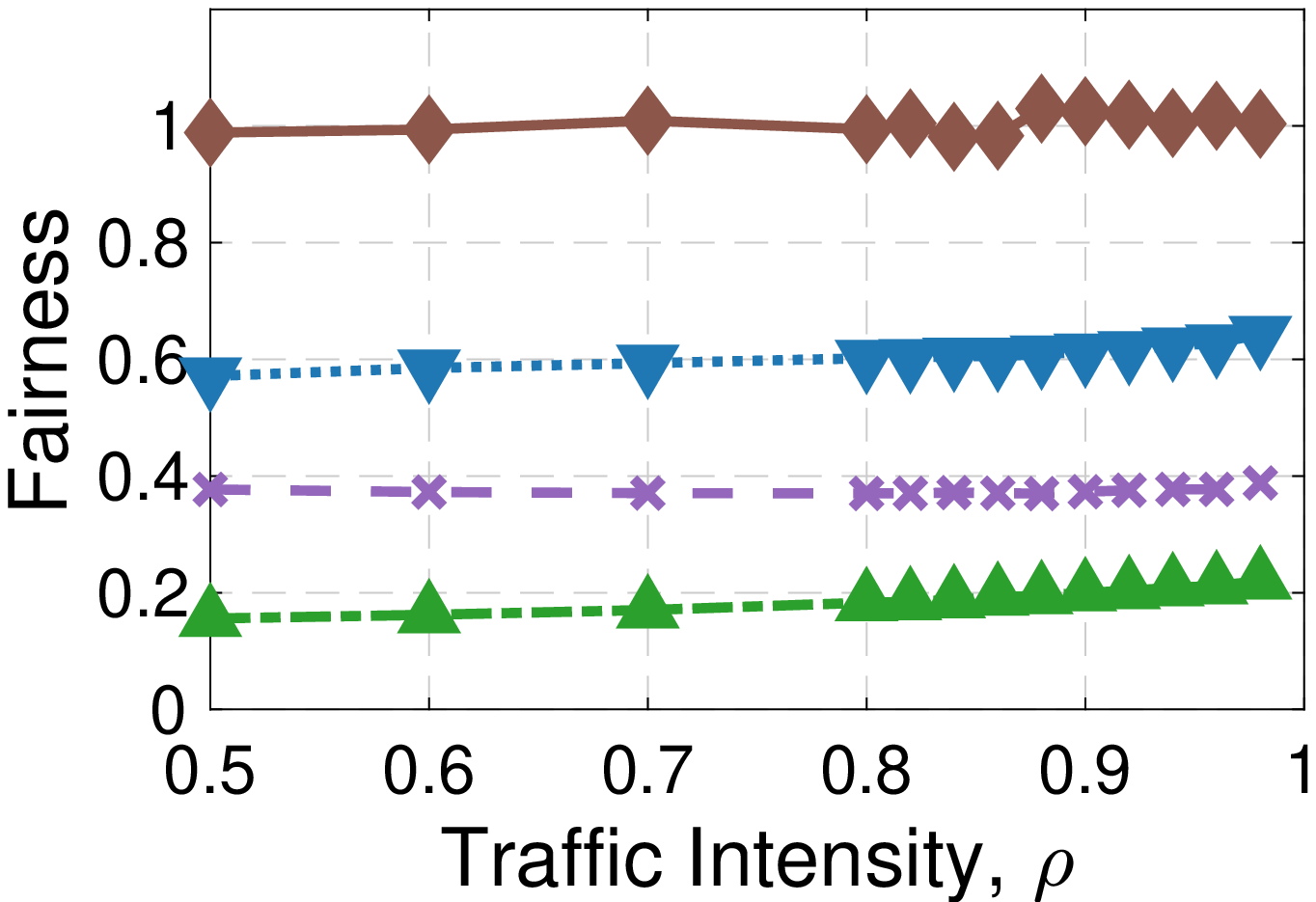}
\label{fig:fairness-equal-rate-ul-vs-dl}
}
\vspace{-0.5\baselineskip}
\caption{Long-term average queue length ratio between (a) FD and HD users, and (b) ULs and DLs with varying traffic intensity, in an HD-FD network with $\numUserFD = \numUserHD = 5$ and equal arrival rates.}
\vspace{-\baselineskip}
\label{fig:fairness-equal-rate}
\end{figure}

Fig.~\ref{fig:fairness-equal-rate}\subref{fig:fairness-equal-rate-fd-vs-hd} plots the fairness between FD and HD users in an HD-FD network with $\numUserFD = \numUserHD = 5$ and varying traffic intensity, $\rho$. It can be observed that {\HGMSRDL} has the worst fairness performance since the DL participating in the channel contention is selected uniformly at random by the AP. When the traffic intensity is low or moderate, {\QCSMA} and {\HGMSLDL} achieve similar fairness of about $0.5$. This is because under equal arrival rates, FD queues are about half the length of the HD queues due to the fact that they are being served about twice as often (i.e., an FD bi-directional transmission can be either activated by the FD UL or DL due to the FD PHY capability). When the traffic intensity is high, both {\HGMSLDL} and {\HGMSLDLE} have increased fairness performance since the longest DL queue will be served more often due to the central DL queue information at the AP. Furthermore, {\HGMSLDLE} outperforms {\HGMSLDL} since, under {\HGMSLDLE}, the AP not only has explicit information of all the DL queues, but also has estimated UL queue lengths that can be used to better assign the access probability distribution $\probAccVec$.

Fig.~\ref{fig:fairness-equal-rate}\subref{fig:fairness-equal-rate-ul-vs-dl} presents the fairness between ULs and DLs with the same network setting. It can be seen that {\QCSMA} has the best fairness performance of around $1$ since all the $2\numUser$ link have equal access probability. The fairness by {\HGMSRDL} between FD and HD users, and between ULs and DLs, are almost identical, and are always the worst among all variants of {\HGMS}. On the other hand, {\HGMSLDLE} still has the best fairness performance among all variants of {\HGMS} by leveraging the information on estimated UL queue lengths.

\begin{figure}[!t]
\centering
\vspace{-0.5\baselineskip}
\subfloat[Traffic intensity $\rho = 0.8$]{
\includegraphics[width=0.45\columnwidth]{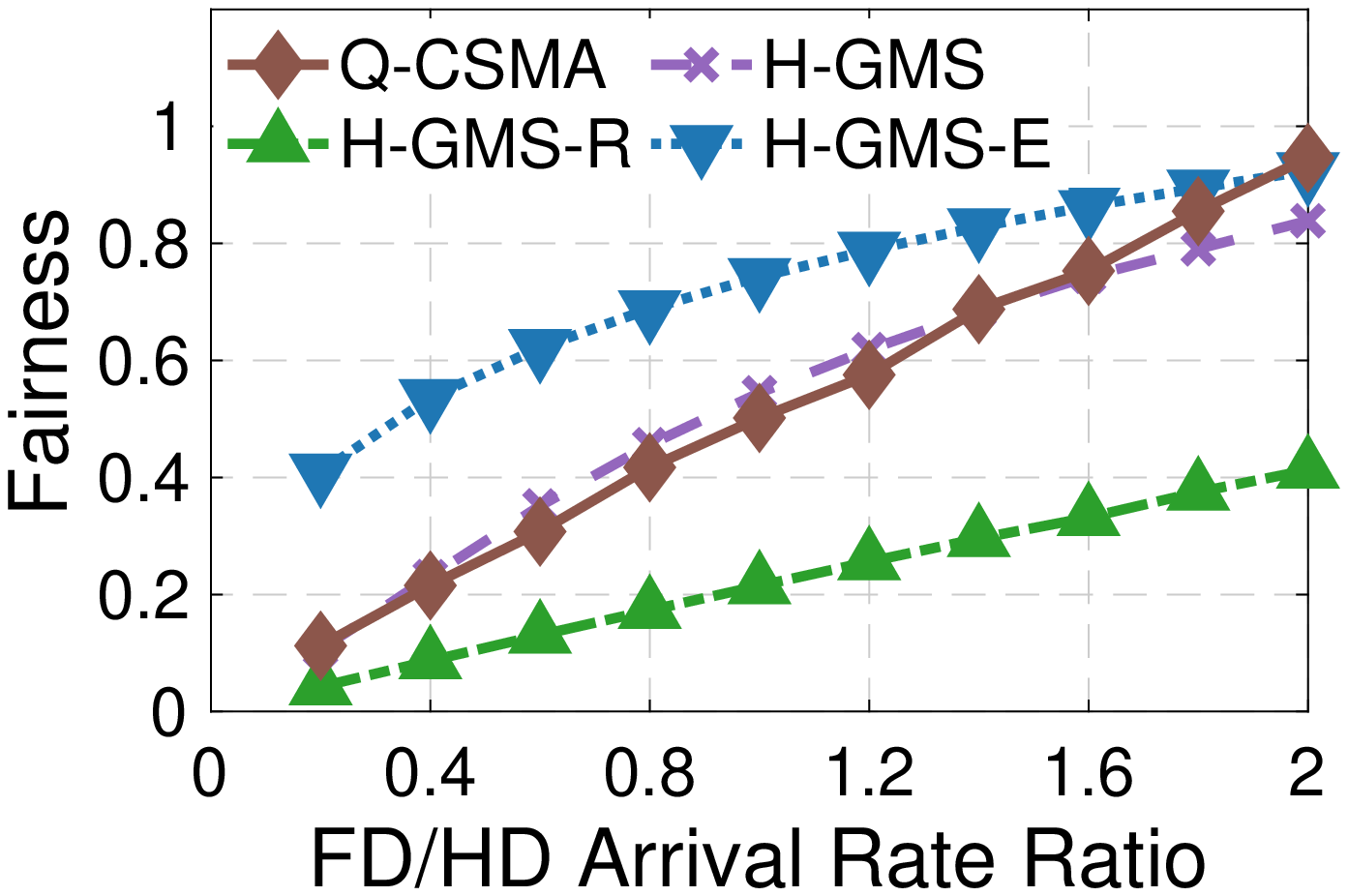}
\label{fig:fairness-diff-rate-fd-vs-hd-load-medium}
}
\subfloat[Traffic intensity $\rho = 0.95$]{
\includegraphics[width=0.45\columnwidth]{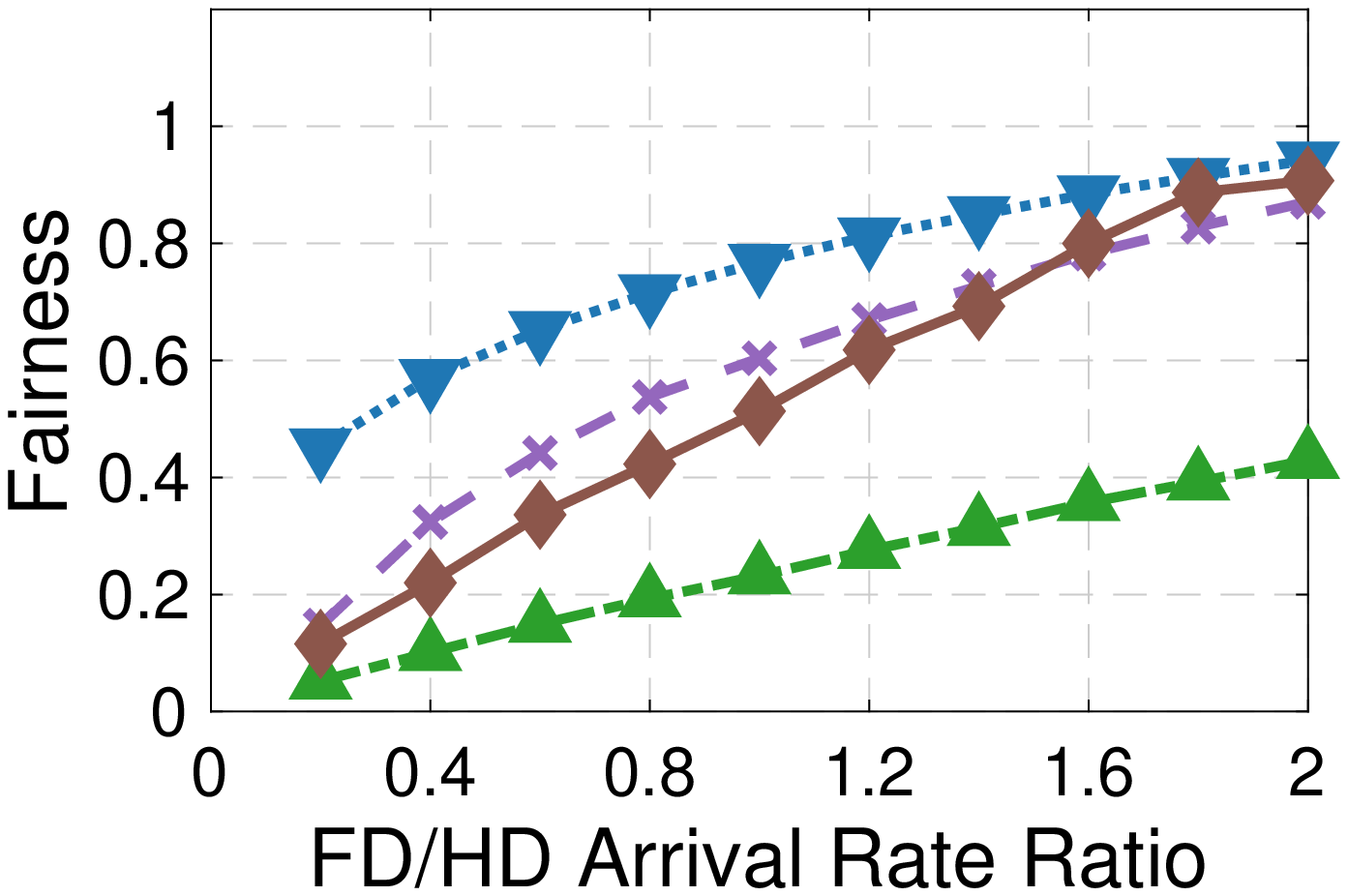}
\label{fig:fairness-diff-rate-fd-vs-hd-load-high}
}
\vspace{-0.5\baselineskip}
\caption{Long-term average queue length ratio between FD and HD users in a heterogeneous HD-FD network with $\numUserFD = \numUserHD = 5$ and varying ratio between FD and HD arrival rates, with (a) moderate ($\rho=0.8$), and (b) high ($\rho=0.95$) traffic intensities.}
\vspace{-\baselineskip}
\label{fig:fairness-diff-rate-fd-vs-hd}
\end{figure}

\subsubsection{Different Arrival Rates}
We also evaluate the fairness under different arrival rates between FD and HD users. Let $\sigma$ be the ratio between the arrival rates on FD and HD links. It is easy to see that if we assign $v_{f} = \sigma/(\sigma\numUserFD + 2\numUserHD)$ and $v_{h} = 1/(\sigma\numUserFD + 2\numUserHD)$, then $\mathbf{v}$ is on the boundary of $\capReg_{\rm HD-FD}$. In this case, we have a capacity region expansion value of $\rateImprov = 1+\sigma\numUserFD/(\sigma\numUserFD+2\numUserHD)$, which depends on both $\numUserFD$ and $\sigma$ (see Section~\ref{ssec:model-cap-region}).

Fig.~\ref{fig:fairness-diff-rate-fd-vs-hd} plots the fairness between FD and HD users with varying $\sigma$ under moderate ($\rho=0.8$) and high ($\rho=0.95$) traffic intensities on the $x$-axis. It can be observed that as the packet arrival rate at FD users increases, the FD and HD queue lengths are better balanced. When $\sigma=2$, FD and HD users have almost the same average queue length since the FD queues are served twice as often as the HD queues under {\QCSMA}, {\HGMSLDL}, and {\HGMSLDLE}. It is interesting to note that the fairness under {\QCSMA} and {\HGMSLDL} is almost a linear function with respect to the arrival rate ratio, $\sigma$. This is intuitive since, as the FD queues are served about twice as often as the HD queues, increased arrival rates will result in longer queue lengths at the FD users. Moreover, since the FD and HD queues have about the same queue length when $\sigma$ approaches $2$, {\HGMSLDLE} does not further improve the fairness since it generates an access probability distribution that is approximately a uniform distribution.

\subsubsection{Impact of the Number of FD Users, $\numUserFD$}

\begin{figure}[!t]
\centering
\vspace{-0.5\baselineskip}
\subfloat[Traffic intensity $\rho=0.8$]{
\includegraphics[width=0.45\columnwidth]{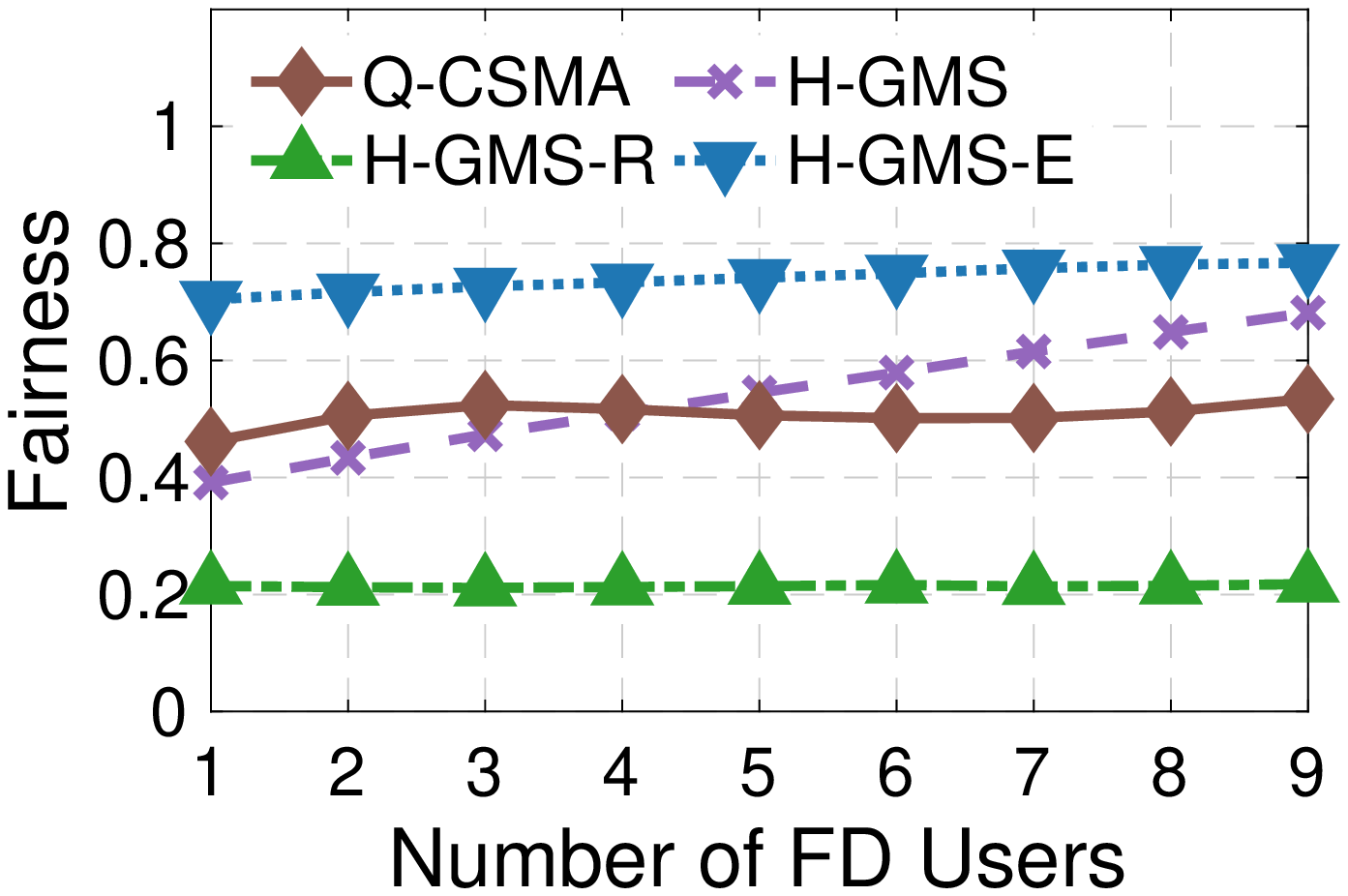}
\label{fig:fairness-diff-nfd-fd-vs-hd-load-medium}
}
\subfloat[Traffic intensity $\rho=0.95$]{
\includegraphics[width=0.45\columnwidth]{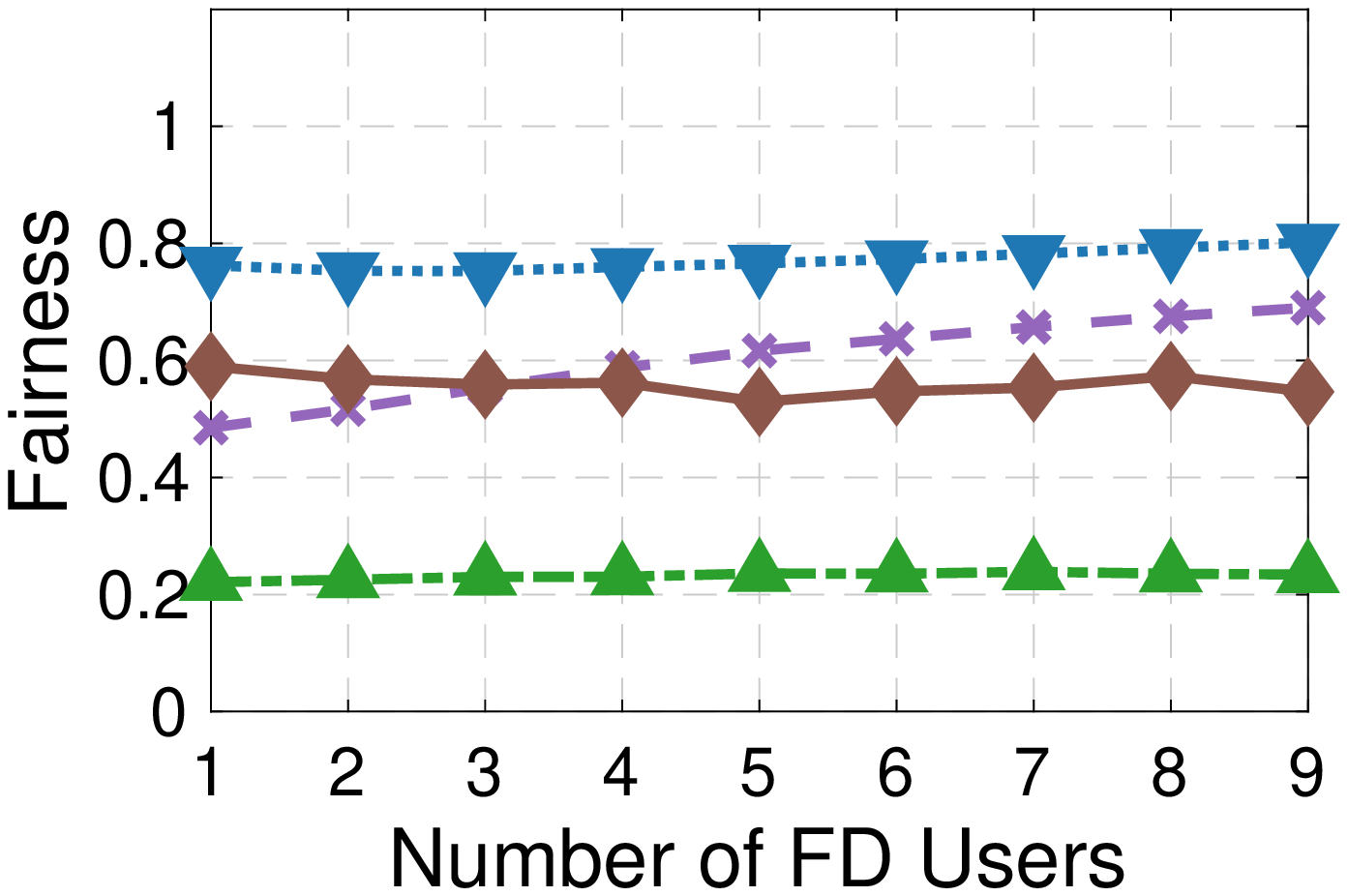}
\label{fig:fairness-diff-nfd-fd-vs-hd-load-high}
}
\vspace{-0.5\baselineskip}
\caption{Long-term average queue length ratio between FD and HD users in a heterogeneous HD-FD network with $\numUserFD = \numUserHD = 5$ and varying $\numUserFD \in \{1,2,\cdots,\numUser-1\}$, with (left) moderate ($\rho=0.8$), and (right) high ($\rho=0.95$) traffic intensities.}
\label{fig:fairness-diff-nfd-fd-vs-hd}
\vspace{-\baselineskip}
\end{figure}

We now evaluate the fairness between FD and HD users with varied number of FD (or equivalently, HD) users under the equal arrival rate model. We vary $\numUserFD \in \{1,2,\cdots,9\}$. Fig.~\ref{fig:fairness-diff-nfd-fd-vs-hd} plots the fairness between FD and HD users under moderate ($\rho=0.8$) and high ($\rho=0.95$) traffic intensities.

As Fig.~\ref{fig:fairness-diff-nfd-fd-vs-hd} suggests, the fairness depends on the number of FD users, $\numUserFD$, only under {\HGMSLDL}. This is because under equal arrival rate, FD users have about half the queue lengths compared with HD users. As $\numUserFD$ increases, the number of HD DLs at the AP (those with relatively larger queue length) decreases and as a result, the AP is very likely to select an HD DL or UL under the {\HGMSLDL} algorithm, resulting in larger average queue length at the FD users. In addition, {\HGMSLDLE} resolves this issue by taking into account the UL queue length estimates. Therefore, the FD users that have smaller queues will be selected with a lower probability so that the longer HD queues will be served at a higher rate. In addition, as $\numUserFD$ increases, {\HGMSLDL} achieves better fairness than that of the classical {\QCSMA} by approximating the {\GMS} (instead of {\MWS} as {\QCSMA} does) in a distributed manner. Moreover, {\HGMSLDLE} has the best fairness performance which is independent of the value of $\numUserFD$.

\subsection{Impact of the Weight Function, $f(x)$}
\label{ssec:sim-diff-tx-prob}

We now evaluate the delay performance of {\HGMS} under different weight functions and compare it to {\QCSMA}. Recall from Theorem~\ref{thm:stability} that {\HGMS} is throughput-optimal for a broad family of weight functions, $f(x)$, and the relationship between $f(\cdot)$ and the transmission probability $\probTX(\cdot)$ is given by {\eqref{eq:tx-prob}}. In particular, we consider the following weight functions:
\begin{itemize}[leftmargin=*,topsep=0pt]
\item $f(x) = \frac{1}{2}\log{(1+x)}$: $\lim_{x\to\infty} \frac{f(x)}{\log{x}} = \frac{1}{2} < 1$;
\item $f(x) = \log{(1+x)}$: $\lim_{x\to\infty} \frac{f(x)}{\log{x}} = 1$;
\item $f(x) = \sqrt{x}$: $\lim_{x\to\infty} \frac{f(x)}{\log{x}} = \infty$ ($\beta = \frac{1}{2}$);
\item $f(x) = x$: $\lim_{x\to\infty} \frac{f(x)}{\log{x}} = \infty$ ($\beta = 1$).
\end{itemize}
\begin{figure}[!t]
\centering
\vspace{-\baselineskip}
\subfloat{
\includegraphics[width=0.3\columnwidth]{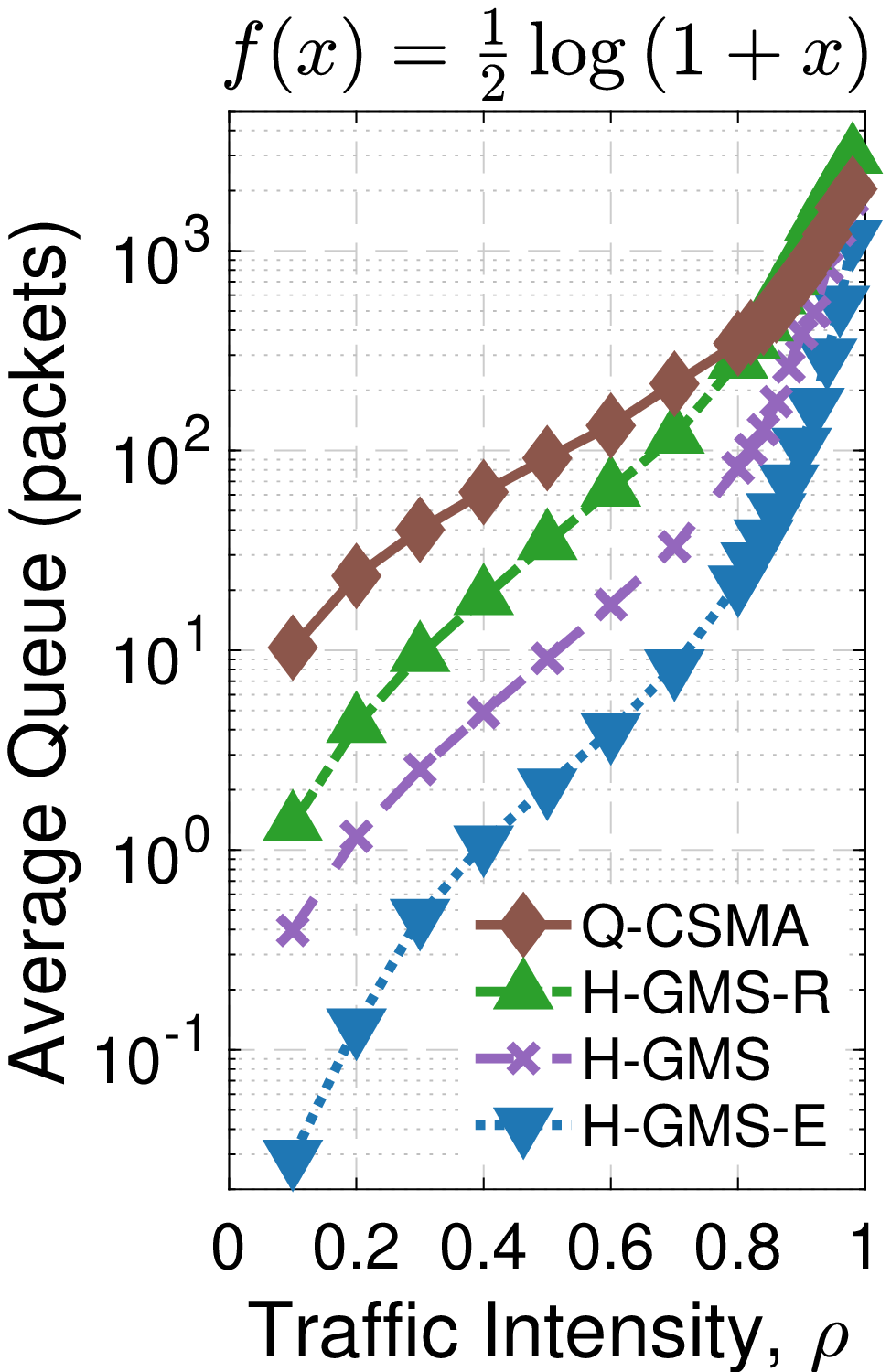}
\label{fig:delay-diff-txprob-func1}
}
\subfloat{
\includegraphics[width=0.3\columnwidth]{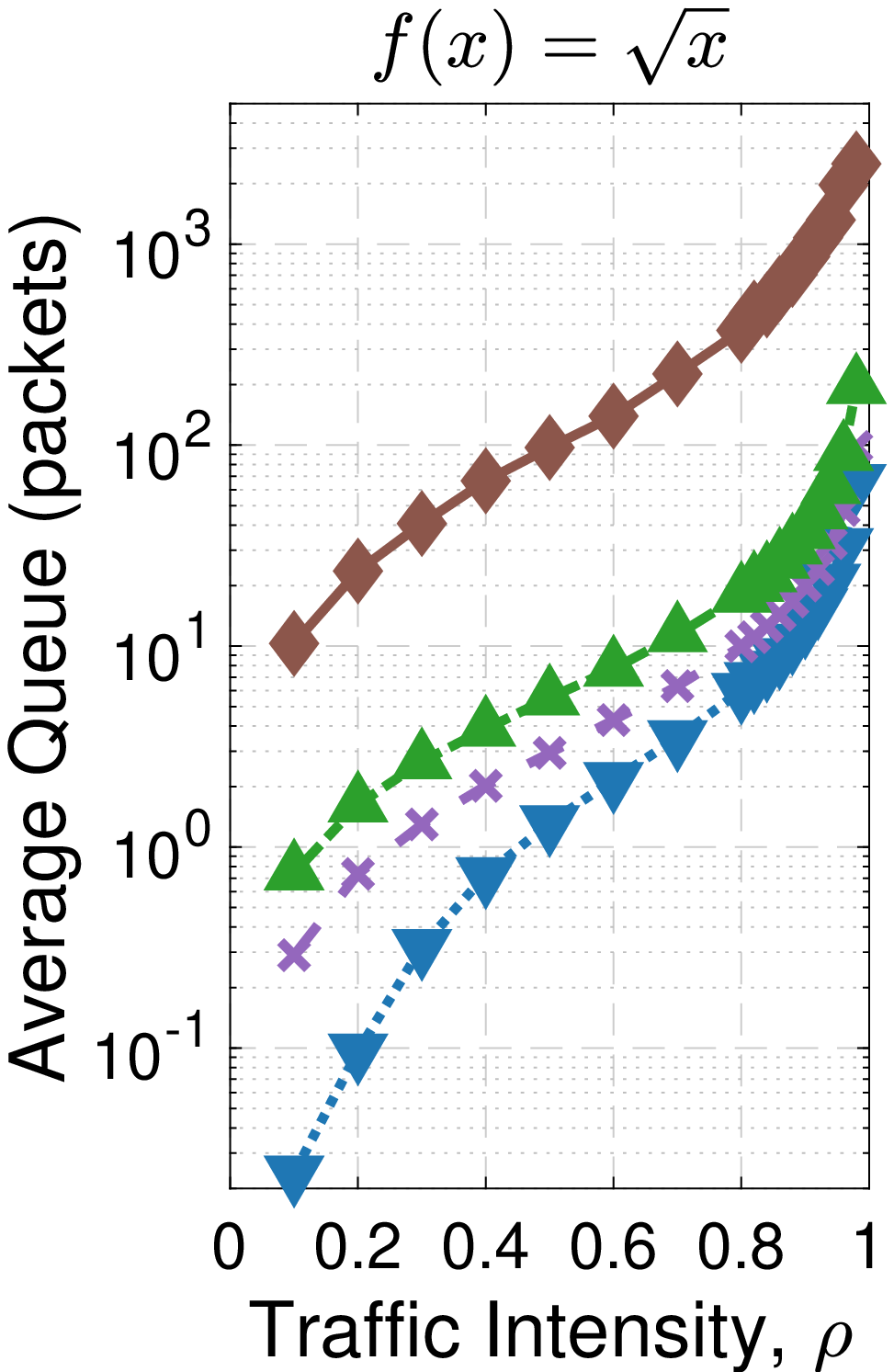}
\label{fig:delay-diff-txprob-func3}
}
\subfloat{
\includegraphics[width=0.3\columnwidth]{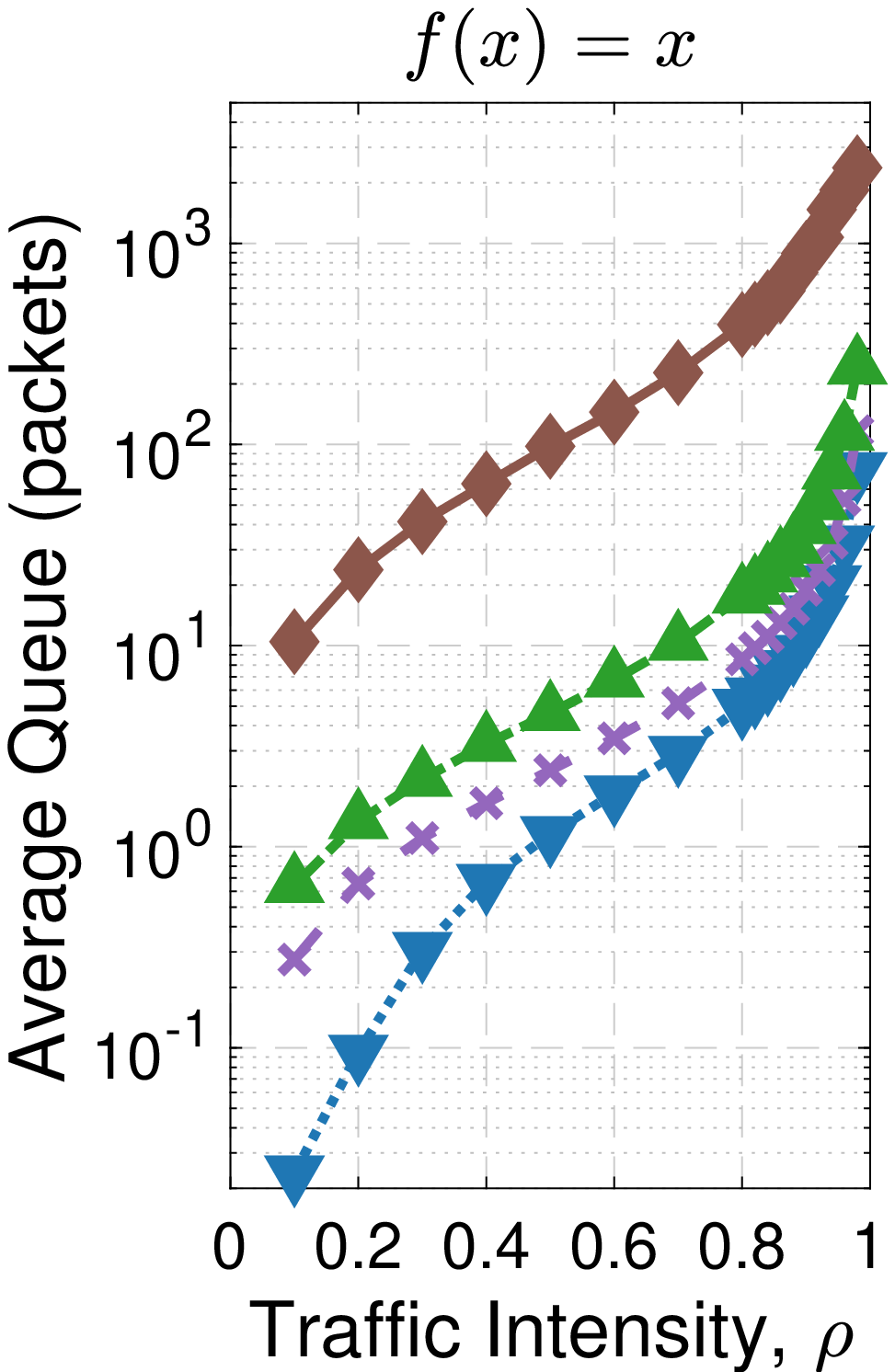}
\label{fig:delay-diff-txprob-func4}
}
\vspace{-0.5\baselineskip}
\caption{Long-term average queue length in a heterogeneous HD-FD network with $\numUserFD = \numUserHD = 5$ and equal arrival rates, with different weight functions and $\probTXLink{\link}(t) = \frac{\exp{(f(\queueLink{\link}(t))}}{1+\exp{(f(\queueLink{\link}(t))}}$. The results in the case with $f(x) = \log{(1+x)}$ are shown in Fig.~\ref{fig:delay-and-lb}(b).}
\label{fig:delay-diff-txprob}
\vspace{-\baselineskip}
\end{figure}

Fig.~\ref{fig:delay-diff-txprob} plots the average queue length with varying traffic intensity in an HD-FD network with $\numUserFD = \numUserHD = 5$ and equal arrival rates, and with different weight functions, $f(x)$, as listed above. {For each considered $f(\cdot)$, we consider all four distributed algorithms listed in Section~\ref{ssec:sim-setup}.} The results in the case with $f(x) = \log{(1+x)}$ are shown in Fig.~\ref{fig:delay-and-lb}\subref{fig:delay-and-lb-n-10-nfd-5}. Table~\ref{tab:delay-diff-txprob-comparison} summarizes the improvements in the delay performance achieved by variants of {\HGMS} compared to {\QCSMA}, with the considered weight functions and moderate ($\rho=0.8$) and extremely high ($\rho=0.98$) traffic intensities.\footnote{The results in the cases with $f(x) = \sqrt{x}$ and $f(x) = x$ are almost identical (see Fig.~\ref{fig:delay-diff-txprob}) and thus omitted in Table~\ref{tab:delay-diff-txprob-comparison}.}

\begin{table}[!t]
\vspace{\baselineskip}
\caption{Improvements in the delay performance achieved by {\HGMS} compared with {\QCSMA} under three different weight functions with different aggressiveness and, with moderate ($\rho=0.8$) and extremely high ($\rho=0.98$) traffic intensities.}
\label{tab:delay-diff-txprob-comparison}
\vspace{-0.5\baselineskip}
\scriptsize
\begin{center}
\renewcommand{\arraystretch}{1}
\begin{tabular}{>{\centering}m{1in}cccccc}
\toprule
Weight Function, $f(x)$ & \multicolumn{2}{c}{$\frac{1}{2}\log{(1+x)}$} & \multicolumn{2}{c}{$\log{(1+x)}$} & \multicolumn{2}{c}{$x$} \\
\midrule
Traffic Intensity, $\rho$ & $0.8$ & $0.98$ & $0.8$ & $0.98$ & $0.8$ & $0.98$ \\
\hline \vspace{1pt}
$\frac{\queueAvg_{\QCSMA}}{\queueAvg_{\HGMSRDL}}$ & $1.2$ & $0.7$ & $14.4$ & $8.5$ & $22.3$ & $9.8$ \\
\vspace{3pt}
$\frac{\queueAvg_{\QCSMA}}{\queueAvg_{\HGMSLDL}}$ & $4.2$ & $1.1$ & $28.4$ & $16.2$ & $46.2$ & $20.4$ \\
\vspace{3pt}
$\frac{\queueAvg_{\QCSMA}}{\queueAvg_{\HGMSLDLE}}$ & $15.8$ & $1.7$ & $52.8$ & $25.4$ & $79.2$ & $31.8$ \\
\bottomrule
\end{tabular}
\renewcommand{\arraystretch}{1}
\vspace{-1.5\baselineskip}
\end{center}
\normalsize
\end{table}

The results show that all the scheduling algorithms are throughput-optimal under different choices of $f(x)$ that satisfy the conditions in Theorem~\ref{thm:stability}. Overall, the delay performance of {\QCSMA} in HD-FD networks has less dependency on $f(x)$ than {\HGMS}, and variants of {\HGMS} (especially {\HGMSLDL} and {\HGMSLDLE}) achieve significantly improved delay performance. Moreover, the delay improvement achieved by {\HGMS} over the classical {\QCSMA} becomes more significant with a ``more aggressive" weight function. For example, {\HGMSLDL} with a sublinear/linear weight function ($f(x)=x^{\beta}$ with $\beta \in \{\frac{1}{2},1\}$) achieves $10$--$20\times$ better delay than  with a logarithmic weight function $f(x)=\frac{1}{2}\log{(1+x)}$. This highlights the importance of the selection of $f(x)$ in the design of {\HGMS}.


%% file: tex/proof_OLoP.tex
The proof is based on the structural properties of the interference graph of the heterogeneous HD-FD network. The interference (or conflict) graph is defined as $G_{\textrm{I}} = (\setNode_{\textrm{I}},\setLink_{\textrm{I}})$, where  $\setNode_{I}$ is the set of network links, and there is an edge between link $\link_{i}$ and link $\link_{j}$ if they interfere with each other. Clearly, the interference graph of a collocated all-HD network is a clique. For the collocated  HD-FD network, $\setNode_{\textrm{I}}=\{v_{i}^{j}, i \in \setNode, j \in \setULDL\}$, where $v_{i}^{j}$ corresponds to link $j$ (UL or DL) of user $i$. Since a pair of FD UL and DL can be simultaneously activated, $G_{\textrm{I}}$ is a complete graph with $\numUserFD$ edges missing, each of which has endpoints $(v_{i}^{\textrm{u}},v_{i}^{\textrm{d}})$ for $\forall i \in \setUserFD$. 

It has been shown in~\cite{dimakis2006sufficient} that the Greedy Maximal Scheduling ({\GMS}) is throughput-optimal if the interference graph $G_{\textrm{I}}$ satisfies the so called \emph{Overall Local Pooling (OLoP)} condition. 
We use the following definition and result from~\cite{birand2012analyzing}.
\begin{definition}[Co-strongly perfect graph]\label{def:co-strongly-perfect-graph}
A graph $G$ is co-strongly perfect, if and only if $G$ contains a clique that intersects every maximal independent set in $G$. 
\end{definition}
\begin{proposition}[\hspace{1sp}{\cite[Definitions~2.2, 2.3, and 5.1]{birand2012analyzing}}]
\label{prop:co-stp-olop}
Every graph that is co-strongly perfect satisfies OLoP.
\end{proposition}
We now prove Proposition~\ref{prop:GMS}. To show that $G_{\textrm{I}}$ is co-strongly perfect, if suffices to find a clique contained in $G_{\textrm{I}}$ that intersects every maximal independent set in $G_{\textrm{I}}$. Recall that $G_{\textrm{I}}$ is a complete graph with $\numUserFD$ edges missing. Let $K = \{\nodeUL{n}, n \in \setUser\} \cup \{\nodeDL{m}, m \in \setUserHD\} \subseteq \setNode_{\textrm{I}}$ with $|K| = \numUserFD + 2\numUserHD$. It is easy to see that the induced graph $G(K)$ on $K$ is a clique. In addition, note that the maximal independent set in $G_{\textrm{I}}$ can be (i) $\{\nodeUL{m}\}$ or $\{\nodeDL{m}\}$ for some $m \in \setUserHD$, or (ii) $\{\nodeUL{n},\nodeDL{n}\}$ for some $n \in \setUserFD$ (there are a total number of $(\numUserFD+2\numUserHD)$ such maximal independent sets). Thus, $G(K)$ intersects with every maximal independent set in $G_{\textrm{I}}$, which implies that $G_{\textrm{I}}$ is co-strongly perfect and satisfies OLoP. Hence, the centralized {\GMS} algorithm (described in Algorithm~\ref{alg:GMS}) is throughput-optimal in \emph{any} collocated heterogeneous HD-FD network.


%% file: tex/proof_steady_state.tex
Under fixed $\queueVec(t)=\queueVec$, {\eqref{eq:DTMC-trans-prob}} is the transition probability matrix of the discrete time Markov chain $Y^{\queueVec}$. Recall that the state space of $Y(t)$ is $S_{Y}=\{0,1,\cdots,\numUser,\optDL\}$, where $\optDL=\arg\max_{i\in\setUser}\queueDL{i}$. The detailed balance equations for $Y^{\queueVec}$ are:
\begin{align}
\label{eq:DTMC-DBE}
\pi^{\queueVec}(0) \cdot \transProb{0}{i} & = \pi^{\queueVec}(i) \cdot \transProb{i}{0},\ \forall i \in S_{Y} \setminus \{0\}.
\end{align}
Recall that the transmission probability is given by {\eqref{eq:tx-prob}}; from {\eqref{eq:DTMC-trans-prob}} and {\eqref{eq:DTMC-DBE}}, we have
\begin{align*}
\pi^{\queueVec}(i) & = \probAccLink{i} \frac{\probTXUL{i}}{\probBarTXUL{i}} \cdot \pi^{\queueVec}(0) \\
& = \probAccLink{i} \exp{(f(\queueUL{i}))} \cdot \pi^{\queueVec}(0),\ \forall i \in S_{Y} \setminus \{0,\optDL\}, \\
\pi^{\queueVec}(\optDL) & = \probAccLink{\textrm{AP}} \frac{\probTXDL{\optDL}}{\probBarTXDL{\optDL}} \cdot \pi^{\queueVec}(0) = \probAccLink{\textrm{AP}}\exp{(f(\queueDL{\optDL}))} \cdot \pi^{\queueVec}(0).
\end{align*}
Normalizing $\sum_{s \in S_{Y}}\pi^{\queueVec}(s)=1$ yields the steady-state distribution {\eqref{eq:DTMC-steady-state}} in Lemma~\ref{lem:DTMC-steady-state}, in which
\begin{align*}
Z = 1 + \littlesum_{i \in S_{Y} \setminus \{0,\optDL\}} \probAccLink{i} \exp{(f(\queueUL{i}))} + \probAccLink{\textrm{AP}} \exp{(f(\queueDL{\optDL}))}.
\end{align*}


%% file: tex/proof_fluid_limit_eq.tex
 Equations {\eqref{fluid1}}--{\eqref{fluid2}} hold for any scheduling algorithm and their proof is standard. Equation {\eqref{fluid1}} is obtained by taking the limit in {\eqref{queue-dynamics}}. Equation {\eqref{fluid2}} is by applying the Strong Law of Large Numbers to the arrival process. Further, by the Lipschitz continuity of $s$, the derivative of $s_i^j$ (denoted by $\mu_i^j(t)$)  exists at any regular point $t$ (almost everywhere) and is bounded by its Lipschitz constant (less than one). Equations {\eqref{fluid4}}--{\eqref{fluid6}} are specific to {\HGMS}, and we prove them below.
%
%
We consider two cases depending on the choice of the weight function $f(\cdot)$.

\noindent\textbf{Case 1}: $\lim_{x \to \infty} f(x)/\log{x} = b \in (0,1)$.

Recall from Section~\ref{sec:mainproof} that Markov chain $\{Y^{\queueVec(t)}(s)\}_{s\geq t}$ denotes the dynamics of $Y(s)$, assuming a fixed $\queueVec(s)=\queueVec(t)$ for all $s \geq t$. 
%
Consider a fluid sample path under the {\HGMS} algorithm. Suppose $\mathbf{q}(t) \neq \mathbf{0}$ at a regular point $t$. By Lipschitz continuity, we can find a short interval $(t,t+\epsilon)$, such that $\mathbf{q}(\tau)$ is approximately constant ($\approx \mathbf{q}$) for $\forall \tau \in (t,t+\epsilon)$, its actual change being of order $\epsilon$ for non-zero queues. This implies that for $r$ large enough, all the queues with non-zero fluid limit $q_{i}^{j}>0$ are of size $\mathcal{O}(q_{i}^{j} r)$ in the original process, while all the queues with zero fluid limit are of size $o(r)$ in the original process. Therefore, taking the limit $r \to \infty$ in {\eqref{eq:DTMC-steady-state}}, it follows that for any $\queueVec(\tau)$, $\tau \in (rt,rt+r\epsilon)$, $\pi^{\queueVec(\tau)}\to \widetilde{\pi}^{\mathbf{q}}$,
where
\begin{align*}
\widetilde{\pi}^{\mathbf{q}}(i) & = \probAccLink{i}(q_{i}^{\textrm{u}})^{b} / \widetilde{Z}^{\mathbf{q}},\ i \in S_{Y} \setminus \{0, \optDL\}, \\
\widetilde{\pi}^{\mathbf{q}}(\optDL) & = \probAccLink{\textrm{AP}}(q_{\optDL}^{\rm d})^{b} / \widetilde{Z}^{\mathbf{q}},\ 
\widetilde{\pi}^{\mathbf{q}}(0) = 1/\widetilde{Z}^{\mathbf{q}},
\end{align*}
 $\widetilde{Z}^{\mathbf{q}} = 1+\sum_{i'=1}^{\numUser} \probAccLink{i'}(q_{i'}^{\textrm{u}})^{b} + \probAccLink{\textrm{AP}}(q_{\optDL}^{\textrm{d}})^{b}$, and the probabilities are zero for queues which are 0 at the fluid limit.
This shows that, with high probability, a queue with a zero fluid limit \emph{cannot} initiate transmission in steady-state. Hence, in equilibrium, the Markov chain never activates an HD link with empty fluid limit queue or an FD link with empty (both) UL and DL queues. Next, we argue that at any $\tau \in (rt,rt+r\epsilon)$, the Markov chain $Y^{\queueVec(\tau)}$ is at its equilibrium distribution $\pi^{\queueVec(\tau)}= \widetilde{\pi}^{\mathbf{q}}$ as $r \to \infty$.
\begin{proposition}[Mixing time of Markov chain $Y^{\queueVec}$]
Let $\nu_{\tau}$ and $\pi$ denote the instantaneous and the equilibrium distribution of Markov chain $\{Y^{\queueVec}(\tau)\}_{\tau \geq 1}$, respectively. Given $0< \zeta <1$, the mixing time is defined as
\begin{align*}
T_{\textrm{mix}}(\zeta) := \inf \big\{\tau \geq 1: \sup_{s \in S_{Y}} \left|\nu_{\tau}(s)-\pi(s)\right| \leq \zeta \big\}.
\end{align*}
Let $\probAccLink{\textrm{min}} = \min_{i}\{\probAccLink{i}\}$ and $\queue_{\textrm{max}} = \max_{i,j}\{\queue_{i}^{j}\}$. Then
\begin{align*}
T_{\textrm{mix}}(\zeta) \leq  \frac{2 \exp{(f(\queue_{\textrm{max}}))}}{\probAccLink{\textrm{min}}} \cdot \Big[ \log{\Big(\frac{2}{\zeta \probAccLink{\textrm{min}}}\Big)} + f(\queue_{\textrm{max}}) \Big].
\end{align*}
\end{proposition}
\begin{IEEEproof}
The proof follows the application of Raleigh Theorem to characterize the second largest eigenvalue modulus (SLEM) of the transition probability matrix of the Markov chain $Y^{\queueVec}$. The analysis is similar to~\cite[Lemma~5]{ghaderi2013impact} with minor modifications and is omitted.
\end{IEEEproof}

Hence for the Markov chain $Y^{\queueVec(rt)}$, the mixing time is $T_{\textrm{mix}}(1/r) = \mathcal{O}(r^{b}\log{r})$. This shows that for $b<1$, the mixing time is sub-linear in $r$ which completely vanishes when taking the average at the fluid scale, i.e.,
\begin{align*}
\frac{1}{\epsilon} \left(s_{i}^{\textrm{u}}(t+\epsilon)-s_{i}^{\textrm{u}}(t) \right)
& \approx \frac{1}{r\epsilon}\littlesum_{\tau=rt}^{rt+ r\epsilon} \mathds{1}(Y^{\queueVec(tr)}(\tau)=i) \\
& \to  \widetilde{\pi}^{\mathbf{q}}(i),\ \mbox{as } r \to \infty
\end{align*}
where the second convergence is almost surely by the Ergodic Theorem. This indicates that $\mu_{i}^{\textrm{u}}(t) = \widetilde{\pi}^{\mathbf{q}}(i)$, $\forall i \in \setUser$, and similarly, $\mu_{\optDL}^{\textrm{d}}(t) = \widetilde{\pi}^{\mathbf{q}}(\optDL)$. This implies that $\mu_{i}^{j}(t) = 0$ for $i \in \setUserHD$ if $q_{i}^{j}(t)=0$, $j \in \setULDL$, which establishes {\eqref{fluid4}}. Similarly, considering the coordination among the activation of a pair of FD UL and DL, $\mu_{i}^{j}(t) = 0$, $j \in \setULDL$, $i \in \setUserFD$, if $\max(q_{i}^{\textrm{u}}(t), q_{i}^{\textrm{d}}(t)) = 0$, giving  {\eqref{fluid5}}. Also, once an FD UL (or DL) queue initiates the transmission at rate $\mu_{i}^{\textrm{u}}$ (or $\mu_{i}^{\textrm{d}}$), the corresponding DL (or UL), if nonzero, can follow the same rate. This establishes {\eqref{fluid5b}}. Finally, {\eqref{fluid6}} comes from the fact that if $\mathbf{q}(t) \neq \mathbf{0}$, no queue that is empty in the fluid limit can initiate transmission at a positive rate and thus the  non-empty queues transmit at the maximum sum rate of $1$.

\noindent\textbf{Case 2}: $\lim_{x \to \infty} f(x)/\log{x} = b > 1$.

The analysis in this case is similar to the analysis of aggressive CSMA algorithms in~\cite{feuillet2010random,ghaderi2014queue}. Suppose that $\mu_{i}^{j}(t) > 0$ and $q_{i}^{j}(t) > 0$ for some initiator queue. This implies that for some $\epsilon$, $X_{i}^{j}(\tau) = 1$ for $\forall \tau \in (rt, r(t+\epsilon))$, and $\queue_{i}^{j}(\tau) \geq (q_{i}^{j}(t)-\epsilon)r$. The probability that this queue releases the channel after one packet transmission is less than $({\queue_{i}^{j}(\tau)})^{-b}$ which is $\mathcal{O}(r^{-b})$ for $b>1$. The probability that the link releases the channel during any time $\tau \in (rt, r(t+\epsilon))$ is thus less than $\sum_{\tau=rt}^{r(t+\epsilon)} r^{-b}$ which is $\mathcal{O}(r^{1-b})$ which goes to $0$ as $r \to \infty$. This shows that at the fluid limit, if $\mu_{i}^{j}(t)>0$ and $q_{i}^{j}(t)>0$, then $\mu_{i}^{j}(t)=1$. Hence, any positive period of transmission, no matter how short, must be followed by full transmission at rate $1$ until the queue has drained on the fluid scale. Furthermore, when the queue hits zero, another non-zero queue will capture the channel without any capture delay (the proof is similar to that of~\cite[Lemmas~8 and~9]{ghaderi2014queue}). 

This implies that in the heterogeneous HD-FD network, whenever the initiator queue $q_{i}^{j}$ belongs to an HD user $i$, the queue $q_{i}^{j}$ drains at full rate $1$ until it becomes empty at fluid scale. Whenever the initiator queue $q_{i}^{j}$ ($j \in \setULDL$) belongs to an FD user $i$, both its UL and DL queues $q_{i}^{\textrm{u}}$ and $q_{i}^{\textrm{d}}$ can drain at the maximum rate of $1$, until the initiator queue hits zero, at which point both queues release the channel (due to the coordination among a pair of FD UL and DL in Algorithm~\ref{alg:HGMS}). Whenever an HD or FD user releases the channel, another HD or FD user will capture the channel immediately and start transmission at full rate. The choice of which user and which queue captures the channel is randomized over non-zero queues according to access probabilities $\probAccVec$ and whether $\optDL(t) \in \setUserFD$ or $\optDL(t) \in \setUserHD$. Nevertheless, as long as $\mathbf{q}(t) \neq \zeroVec$, an HD link with non-zero queue or an FD link with at least one non-zero queue (either UL or DL) will be activated at full rate. This shows that the fluid limits still satisfy {\eqref{fluid4}}--{\eqref{fluid6}}.

Hence, the fluid limit equations hold for both cases.


%% file: tex/proof_delay_lb_hgms.tex
Recall from Section~\ref{sec:delay}, $\queueAvg = \frac{\queueSum_{\setLinkMax}+\queueSum_{\setLinkMin}}{2\numUser} \geq \frac{\queueSum_{\setLinkMax}}{2\numUser}$. Since the queueing dynamics in $\setLinkMax$ and $\setLinkMax$ are \emph{not} independent due to the existence of FD users, we wish to find a lower bound on $\queueSum_{\setLinkMax}$. Denote $\queue_{\link,\setLinkMax}$ as the queue length of link $\link$ at an arbitrary epoch during a \emph{non-serving} interval for the clique $\setLinkMax$. Denote $\queueSum_{\widetilde{\setLinkMax}} = \littlesum_{\link \in \setLinkMax} \myE{\queue_{\link,\setLinkMax}}$. From the workload decomposition rule~\cite{boxma1989workloads} applied to a discrete time GI/G/1 system in clique $\setLinkMax$, we have~\cite{bouman2014delay}
\begin{align}
\label{eq:workload-decomp}
\queueSum_{\setLinkMax} = \littlesum\nolimits_{\link \in \setLinkMax} \myE{\queue_{\link}} & = \queueSum_{\setLinkMax}^{\textrm{LB}} + \littlesum\nolimits_{\link \in \setLinkMax} \myE{\queue_{\link,\setLinkMax}} \nonumber \\
& = \queueSum_{\setLinkMax}^{\textrm{LB}} + \queueSum_{\widetilde{\setLinkMax}}.
\end{align}
Note that $\queueSum_{\setLinkMax}^{\textrm{LB}}$ and $\queueSum_{\widetilde{\setLinkMax}}$ are both non-negative, and an immediate lower bound on $\queueAvg$ is obtained by
\begin{align}
\label{eq:proof-delay-lb-fundamental}
\queueAvg & = \frac{\littlesum_{\link \in \setLink} \myE{\queueLink{\link}}}{2\numUser} = \frac{\queueSum_{\setLink}}{2\numUser} \geq \frac{\queueSum_{\setLinkMax}}{2\numUser} \geq \frac{\queueSum_{\setLinkMax}^{\textrm{LB}}}{2\numUser}.
\end{align}

A key observation to derive the improved lower bound is that assuming the system is stable, in each time slot, the probability that link $\link$ transitions from idle state to active state (i.e., \emph{link $\link$ is activated}) equals the probability it transitions from active state back to idle state (i.e., \emph{link $\link$ is deactivated}). Therefore, for any link $\link \in \setLinkMax$,
\begin{align}
\label{eq:delay-observation}
\myProb{\link\ \textrm{is activated}} & = \myProb{\link\ \textrm{is deactivated}}.
\end{align}
Let $\partial_{\link}$ denote the set of conflicting links of $\link$ including link $\link$ itself {and recall that the access probability $\probAccVec$ is \emph{fixed} under {\HGMSLDL} and {\HGMSRDL}}. For $\forall \link \in \setLinkMax$,
\begin{align}
\label{eq:delay-prob-activate}
& \myProb{\link\ \textrm{is activated}} = \myE{\probAccLink{\link} \cdot \probTX(\queueLink{\link}) \cdot \mathds{1}(\schedLink{\link'}(t) = 0, \forall \link' \in \partial_{\link})} \nonumber \\
& \quad \leq \myE{\probAccLink{\link} \cdot \probTX(\queueLink{\link}) \cdot \mathds{1}(\schedLink{\link'}(t) = 0, \forall \link' \in \setLinkMax)} \nonumber \\
& \quad = \probAccLink{\link} \myE{\probTX(\queueLink{\link,\setLinkMax})} \cdot \myProb{\schedLink{\link'}(t) = 0, \forall \link' \in \setLinkMax} \nonumber \\
& \quad = \probAccLink{\link} \myE{\probTX(\queueLink{\link,\setLinkMax})} \cdot (1 - \littlesum\nolimits_{\link' \in \setLinkMax} \myProb{\schedLink{\link'}(t) = 1}) \nonumber \\
& \quad = \probAccLink{\link} \myE{\probTX(\queueLink{\link,\setLinkMax})} \cdot (1 - \littlesum\nolimits_{\link' \in \setLinkMax}\pi_{\link'}),
\end{align}
where $\pi_{\link'}$ is the steady state probability of link $\link'$ being active. Similarly,
\begin{align}
\myProb{\link\ \textrm{is deactivated}}
\label{eq:delay-prob-deactivate}
& = \myE{(1-\probTX(\queueLink{\link})) \cdot \mathds{1}(\schedLink{\link}(t) = 1} \nonumber \\
& = \left(1 - \myE{\probTX(\queueLink{\link})} \right) \cdot \pi_{\link}.
\end{align}
Plugging {\eqref{eq:delay-prob-activate}} and {\eqref{eq:delay-prob-deactivate}} into {\eqref{eq:delay-observation}} yields
\begin{align}
\label{eq:delay-queue-inequality-link}
& \probAccLink{\link} \myE{\probTX(\queueLink{\link,\setLinkMax})} \cdot (1 - \littlesum\nolimits_{\link' \in \setLinkMax}\pi_{\link'}) \geq \left(1 - \myE{\probTX(\queueLink{\link})} \right) \cdot \pi_{\link} \nonumber \\
& \Leftrightarrow \frac{\myE{\probTX(\queueLink{\link,\setLinkMax})}}{1-\myE{\probTX(\queueLink{\link})}}
\geq \frac{\pi_{\link}/\probAccLink{\link}}{1 - \littlesum_{\link' \in \setLinkMax}\pi_{\link'}}
\geq \frac{\arrRateLink{\link}/\probAccLink{\link}}{1-\arrRateSum{\setLinkMax}},
\end{align}
where the last inequality comes from the fact that in steady state, $\arrRateLink{\link} \leq \pi_{\link}$ for $\forall \link \in \setLinkMax$. Recall the definitions of $\arrRateLink{\textrm{min}}$ and $\probAccLink{\textrm{max}}$ from Proposition~\ref{prop:delay-lb-hgms},
it is easy to see that $\min_{\link \in \setLinkMax} \arrRateLink{\link} \geq \arrRateLink{\textrm{min}}$ and $\max_{\link \in \setLinkMax} \probAccLink{\link} \leq \probAccLink{\textrm{max}}$ (under both {\HGMSLDL} and {\HGMSRDL}.
Applying {\eqref{eq:delay-queue-inequality-link}} to all $\link \in \setLinkMax$, we obtain
\begin{align}
\label{eq:jenson-1}
& \littlesum_{\link \in \setLinkMax} \myE{\probTX(\queueLink{\link,\setLinkMax})} 
\geq \frac{1}{1-\arrRateSum{\setLinkMax}} \cdot \littlesum_{\link \in \setLinkMax} \frac{\arrRateLink{\link}}{\probAccLink{\link}} \left(1-\myE{\probTX(\queueLink{\link})}\right) \nonumber \\
& \geq \frac{1}{1-\arrRateSum{\setLinkMax}} \cdot \frac{\min_{\link \in \setLinkMax} \arrRateLink{\link}}{\max_{\link \in \setLinkMax} \probAccLink{\link}} \cdot \littlesum_{\link \in \setLinkMax} \left(1-\myE{\probTX(\queueLink{\link})}\right) \nonumber \\
& \geq \frac{1}{1-\arrRateSum{\setLinkMax}} \cdot \frac{\arrRateLink{\textrm{min}}}{\probAccLink{\textrm{max}}} \cdot \littlesum_{\link \in \setLinkMax} \left(1-\myE{\probTX(\queueLink{\link})}\right) \nonumber \\
& \geq \frac{1}{1-\arrRateSum{\setLinkMax}} \cdot \frac{\arrRateLink{\textrm{min}}}{\probAccLink{\textrm{max}}} \cdot |\setLinkMax| \cdot \Big( 1 - \frac{\littlesum_{\link \in \setLinkMax} \myE{\probTX(\queueLink{\link})}}{|\setLinkMax|} \Big) \nonumber \\
& \geq \frac{1}{1-\arrRateSum{\setLinkMax}} \cdot \frac{\arrRateLink{\textrm{min}}}{\probAccLink{\textrm{max}}} \cdot |\setLinkMax| \cdot \Big( 1 - \probTX \Big( \frac{\queueSum_{\setLinkMax}}{|\setLinkMax|} \Big) \Big),
\end{align}
where the last inequality comes from applying Jensen's inequality to the concave increasing function $\probTX(\cdot)$, i.e., 
\begin{align*}
\frac{\littlesum_{\link \in \setLinkMax} \myE{\probTX(\queueLink{\link})}}{|\setLinkMax|} & \leq \probTX \Big( \frac{\queueSum_{\setLinkMax}}{|\setLinkMax|} \Big).
\end{align*}
In addition, the left-hand-side of {\eqref{eq:jenson-1}} can be upper bounded using Jensen's inequality,
\begin{align}
\label{eq:jenson-2}
\littlesum_{\link \in \setLinkMax} \myE{\probTX(\queueLink{\link,\setLinkMax})}
& \leq |\setLinkMax| \cdot \probTX \Big( \frac{\queueSum_{\widetilde{\setLinkMax}}}{|\setLinkMax|} \Big)
\leq |\setLinkMax| \cdot \probTX \Big( \frac{\queueSum_{\setLinkMax}}{|\setLinkMax|} \Big),
\end{align}
where the last inequality is due to $\queueSum_{\widetilde{\setLinkMax}} \leq \queueSum_{\setLinkMax}$ (see {\eqref{eq:workload-decomp}}). Putting together {\eqref{eq:jenson-1}} and {\eqref{eq:jenson-2}} yields
\begin{align*}
\queueSum_{\setLinkMax} & \geq |\setLinkMax| \cdot \probTX^{-1} \Big( \frac{ \arrRateLink{\textrm{min}}/\probAccLink{\textrm{max}}}{1 - \arrRateSum{\setLinkMax} + \arrRateLink{\textrm{min}}/\probAccLink{\textrm{max}} } \Big),
\end{align*}
and as a result,
\begin{align}
\label{eq:proof-delay-lb-hgms}
\queueAvg
& \geq \frac{\queueSum_{\setLinkMax}}{2\numUser} 
\geq \Big(1-\frac{\numUserFD}{2\numUser}\Big) \cdot \probTX^{-1} \Big( \frac{ \arrRateLink{\textrm{min}}/\probAccLink{\textrm{max}}}{1 - \arrRateSum{\setLinkMax} + \arrRateLink{\textrm{min}}/\probAccLink{\textrm{max}} } \Big).
\end{align}
Combining {\eqref{eq:proof-delay-lb-fundamental}} and {\eqref{eq:proof-delay-lb-hgms}} leads to {\eqref{eq:delay-lb-hgms}}, completing the proof.


%% file: paper.bbl
\begin{thebibliography}{10}
\providecommand{\url}[1]{#1}
\csname url@samestyle\endcsname
\providecommand{\newblock}{\relax}
\providecommand{\bibinfo}[2]{#2}
\providecommand{\BIBentrySTDinterwordspacing}{\spaceskip=0pt\relax}
\providecommand{\BIBentryALTinterwordstretchfactor}{4}
\providecommand{\BIBentryALTinterwordspacing}{\spaceskip=\fontdimen2\font plus
\BIBentryALTinterwordstretchfactor\fontdimen3\font minus
  \fontdimen4\font\relax}
\providecommand{\BIBforeignlanguage}[2]{{%
\expandafter\ifx\csname l@#1\endcsname\relax
\typeout{** WARNING: IEEEtran.bst: No hyphenation pattern has been}%
\typeout{** loaded for the language `#1'. Using the pattern for}%
\typeout{** the default language instead.}%
\else
\language=\csname l@#1\endcsname
\fi
#2}}
\providecommand{\BIBdecl}{\relax}
\BIBdecl

\bibitem{chen2018hybrid}
T.~Chen, J.~Diakonikolas, J.~Ghaderi, and G.~Zussman, ``Hybrid scheduling in
  heterogeneous half-and full-duplex wireless networks,'' in \emph{Proc. IEEE
  INFOCOM'18}, 2018.

\bibitem{sabharwal2014band}
A.~Sabharwal, P.~Schniter, D.~Guo, D.~W. Bliss, S.~Rangarajan, and R.~Wichman,
  ``In-band full-duplex wireless: Challenges and opportunities,'' \emph{IEEE J.
  Sel. Areas Commun.}, vol.~32, no.~9, pp. 1637--1652, 2014.

\bibitem{duarte2012experiment}
M.~Duarte, C.~Dick, and A.~Sabharwal, ``Experiment-driven characterization of
  full-duplex wireless systems,'' \emph{IEEE Trans. Wireless Commun.}, vol.~11,
  no.~12, pp. 4296--4307, 2012.

\bibitem{bharadia2013full}
D.~Bharadia, E.~McMilin, and S.~Katti, ``Full duplex radios,'' in \emph{Proc.
  ACM SIGCOMM'13}, 2013.

\bibitem{yang2015wideband}
D.~Yang, H.~Y{\"u}ksel, and A.~Molnar, ``A wideband highly integrated and
  widely tunable transceiver for in-band full-duplex communication,''
  \emph{IEEE J. Solid-State Circuits}, vol.~50, no.~5, pp. 1189--1202, 2015.

\bibitem{zhou2017integrated}
J.~Zhou, N.~Reiskarimian, J.~Diakonikolas, T.~Dinc, T.~Chen, G.~Zussman, and
  H.~Krishnaswamy, ``Integrated full duplex radios,'' \emph{IEEE Commun. Mag.},
  vol.~55, no.~4, pp. 142--151, 2017.

\bibitem{krishnaswamy2016spectrum}
H.~Krishnaswamy and G.~Zussman, ``1 {Chip} 2x the bandwidth,'' \emph{IEEE
  Spectrum}, vol.~53, no.~7, pp. 38--54, 2016.

\bibitem{chen2019wideband}
T.~Chen, M.~B. Dastjerdi, J.~Zhou, H.~Krishnaswamy, and G.~Zussman, ``Wideband
  full-duplex wireless via frequency-domain equalization: Design and
  experimentation,'' in \emph{Proc. ACM MobiCom'19 (to appear)}, 2019.

\bibitem{tassiulas1992stability}
L.~Tassiulas and A.~Ephremides, ``Stability properties of constrained queueing
  systems and scheduling policies for maximum throughput in multihop radio
  networks,'' \emph{IEEE Trans. Autom. Control}, vol.~37, no.~12, pp.
  1936--1948, 1992.

\bibitem{dimakis2006sufficient}
A.~Dimakis and J.~Walrand, ``Sufficient conditions for stability of
  longest-queue-first scheduling: Second-order properties using fluid limits,''
  \emph{Adv. Appl. Prob.}, vol.~38, no.~2, pp. 505--521, 2006.

\bibitem{ghaderi2010design}
J.~Ghaderi and R.~Srikant, ``On the design of efficient {CSMA} algorithms for
  wireless networks,'' in \emph{Proc. IEEE CDC'10}, 2010.

\bibitem{ni2012q}
J.~Ni, B.~Tan, and R.~Srikant, ``{Q-CSMA}: Queue-length-based {CSMA/CA}
  algorithms for achieving maximum throughput and low delay in wireless
  networks,'' \emph{IEEE/ACM Trans. Netw.}, vol.~20, no.~3, 2012.

\bibitem{birand2012analyzing}
B.~Birand, M.~Chudnovsky, B.~Ries, P.~Seymour, G.~Zussman, and Y.~Zwols,
  ``Analyzing the performance of greedy maximal scheduling via local pooling
  and graph theory,'' \emph{IEEE/ACM Trans. Netw.}, vol.~20, no.~1, pp.
  163--176, 2012.

\bibitem{chung2015prototyping}
M.~Chung, M.~S. Sim, J.~Kim, D.~K. Kim, and C.-B. Chae, ``Prototyping real-time
  full duplex radios,'' \emph{IEEE Commun. Mag.}, vol.~53, no.~9, 2015.

\bibitem{zhou2014low}
J.~Zhou, A.~Chakrabarti, P.~R. Kinget, and H.~Krishnaswamy, ``Low-noise active
  cancellation of transmitter leakage and transmitter noise in broadband
  wireless receivers for {FDD}/co-existence,'' \emph{IEEE J. Solid-State
  Circuits}, vol.~49, no.~12, pp. 3046--3062, 2014.

\bibitem{flexicon_orbit_arxiv}
T.~Chen, M.~Baraani~Dastjerdi, J.~Zhou, H.~Krishnaswamy, and G.~Zussman,
  ``Open-access full-duplex wireless in the {ORBIT} testbed,'' \emph{arXiv
  preprint arXiv:1801.03069}, 2018.

\bibitem{maravsevic2016capacity}
J.~Mara{\v{s}}evi{\'c} and G.~Zussman, ``On the capacity regions of
  single-channel and multi-channel full-duplex links,'' in \emph{Proc. ACM
  MobiHoc'16}, 2016.

\bibitem{jelenaToN-FD}
J.~Mara\v{s}evi\'{c}, J.~Zhou, H.~Krishnaswamy, Y.~Zhong, and G.~Zussman,
  ``Resource allocation and rate gains in practical full-duplex systems,''
  \emph{IEEE/ACM Trans. Netw.}, vol.~25, no.~1, pp. 292--305, 2017.

\bibitem{li2014rate}
W.~Li, J.~Lilleberg, and K.~Rikkinen, ``On rate region analysis of half-and
  full-duplex {OFDM} communication links,'' \emph{IEEE J. Sel. Areas Commun.},
  vol.~32, no.~9, pp. 1688--1698, Sept. 2014.

\bibitem{goyal2013distributed}
S.~Goyal, P.~Liu, O.~Gurbuz, E.~Erkip, and S.~Panwar, ``A distributed {MAC}
  protocol for full duplex radio,'' in \emph{Proc. Asilomar'13}, 2013.

\bibitem{chen2017probabilistic}
S.-Y. Chen, T.-F. Huang, K.~C.-J. Lin, Y.-W.~P. Hong, and A.~Sabharwal,
  ``Probabilistic medium access control for full-duplex networks with
  half-duplex clients,'' \emph{IEEE Trans. Wireless Commun.}, vol.~16, no.~4,
  2017.

\bibitem{sahai2011pushing}
A.~Sahai, G.~Patel, and A.~Sabharwal, ``Pushing the limits of full-duplex:
  Design and real-time implementation,'' \emph{arXiv preprint: 1107.0607},
  2011.

\bibitem{xie2014does}
X.~Xie and X.~Zhang, ``Does full-duplex double the capacity of wireless
  networks?'' in \emph{Proc. IEEE INFOCOM'14}, 2014.

\bibitem{tang2015duplex}
A.~Tang and X.~Wang, ``A-duplex: Medium access control for efficient
  coexistence between full-duplex and half-duplex communications,'' \emph{IEEE
  Trans. Wireless Commun.}, vol.~14, no.~10, pp. 5871--5885, 2015.

\bibitem{yang2015scheduling}
Y.~Yang and N.~B. Shroff, ``Scheduling in wireless networks with full-duplex
  cut-through transmission,'' in \emph{Proc. IEEE INFOCOM'15}, 2015.

\bibitem{alim2017band}
M.~A. Alim, M.~Kobayashi, S.~Saruwatari, and T.~Watanabe, ``In-band full-duplex
  medium access control design for heterogeneous wireless {LAN},''
  \emph{EURASIP J. Wireless Commun. and Netw.}, vol. 2017, no.~1, p.~83, 2017.

\bibitem{bianchi2000performance}
G.~Bianchi, ``Performance analysis of the ieee 802.11 distributed coordination
  function,'' \emph{IEEE J. Sel. Areas Commun.}, vol.~18, no.~3, 2000.

\bibitem{bouman2011backlog}
N.~Bouman, S.~Borst, J.~van Leeuwaarden, and A.~Proutiere, ``Backlog-based
  random access in wireless networks: Fluid limits and delay issues,'' in
  \emph{Proc. ITC'11}, 2011.

\bibitem{ghaderi2014queue}
J.~Ghaderi, S.~Borst, and P.~Whiting, ``Queue-based random-access algorithms:
  Fluid limits and stability issues,'' \emph{Stochastic Systems}, vol.~4,
  no.~1, pp. 81--156, 2014.

\bibitem{feuillet2010random}
M.~Feuillet, A.~Proutiere, and P.~Robert, ``Random capture algorithms fluid
  limits and stability,'' in \emph{Proc. IEEE ITA'10}, 2010.

\bibitem{whitt1970weak}
W.~Whitt, ``Weak convergence of probability measures on the function space
  $c[0,\infty)$,'' \emph{Ann. of Math. Stat.}, vol.~41, no.~3, pp. 939--944,
  1970.

\bibitem{dai1995positive}
J.~G. Dai, ``On positive {Harris} recurrence of multiclass queueing networks: a
  unified approach via fluid limit models,'' \emph{Ann. Appl. Prob.}, pp.
  49--77, 1995.

\bibitem{gupta2010delay}
G.~R. Gupta and N.~B. Shroff, ``Delay analysis for wireless networks with
  single hop traffic and general interference constraints,'' \emph{IEEE/ACM
  Trans. Netw.}, vol.~18, no.~2, pp. 393--405, 2010.

\bibitem{boxma1989workloads}
O.~J. Boxma, ``Workloads and waiting times in single-server systems with
  multiple customer classes,'' \emph{Queueing Systems}, vol.~5, no. 1-3, 1989.

\bibitem{ghaderi2013impact}
J.~Ghaderi and R.~Srikant, ``The impact of access probabilities on the delay
  performance of {Q-CSMA} algorithms in wireless networks,'' \emph{IEEE/ACM
  Trans. Netw.}, vol.~21, no.~4, pp. 1063--1075, 2013.

\bibitem{bouman2014delay}
N.~Bouman, S.~C. Borst, and J.~S. van Leeuwaarden, ``Delay performance in
  random-access networks,'' \emph{Queueing Systems}, vol.~77, no.~2, 2014.

\end{thebibliography}
